\documentclass[aps,prb,preprint,amsmath,amssymb,superscriptaddress]{revtex4-1}
\usepackage{graphicx}
\usepackage[utf8]{inputenc}
\usepackage{hyperref}
\usepackage{amsmath,amsfonts}
\begin{document}
\title{Magnetotransport signatures of chiral magnetic anomaly in the half-Heusler phase ScPtBi}
\author{Orest Pavlosiuk} 
\affiliation{Institute of Low Temperature and Structure Research, 
Polish Academy of Sciences, ul. Okólna 2, 50-422 Wrocław, Poland} 
\author{Andrzej Jezierski} 
\affiliation{Institute of Molecular Physics, Polish Academy of Sciences, ul. M. Smoluchowskiego 17, 60-179  Poznań, Poland} 
\author{Dariusz Kaczorowski} 
\affiliation{Institute of Low Temperature and Structure Research, 
Polish Academy of Sciences, ul. Okólna 2, 50-422 Wrocław, Poland}
\author{Piotr Wiśniewski} 
\email[Corresponding author: ] {p.wisniewski@intibs.pl}
\affiliation{Institute of Low Temperature and Structure Research, 
Polish Academy of Sciences, ul. Okólna 2, 50-422 Wrocław, Poland}
\begin{abstract} 
Study of the magnetotransport properties of ScPtBi revealed simultaneously: a negative contribution to the longitudinal magnetoresistance, planar Hall effect, and distinct angular narrowing of the longitudinal magnetoresistance -- three hallmarks of chiral magnetic anomaly (pumping of axial charge between Weyl nodes), a distinct property of topological semimetals. 
Electronic structure calculations show that structural defects, such as antisites and vacancies, bring a substantial density of states at the Fermi level of ScPtBi, indicating that it is a semimetal, not a zero-gap semiconductor, as predicted earlier. This is in accord with electrical resistivity in ScPtBi, showing no characteristics of a semiconductor. Moreover, below 0.7\,K we observed an onset of a superconducting transition, with the resistivity disappearing completely below 0.23\,K. 
\end{abstract}
\maketitle
\section{Introduction}
Chiral magnetic anomaly (CMA) is a unique and defining feature of topological semimetals, due to charge pumping between Weyl nodes of opposite chirality.  
Weyl nodes are monopoles of Berry flux and possess characteristic chiral charge (or Chern number, being integral of the Berry flux through a closed surface enclosing the node), equal to $\pm 1$ in a single-Weyl, or to $\pm 2$ in a double-Weyl semimetal~~\cite{Armitage2018}. 

CMA has three main electronic magnetotransport manifestations: negative longitudinal magnetoresistance (LMR)~~\cite{Nielsen1983,Son2013}, the planar Hall effect (PHE), and angular narrowing of negative LMR~~\cite{Xiong2015, Burkov2017}.
All three have never been observed simultaneously in the same material.
Negative LMR has been found in many topological semimetals, for example: TaAs, TaP, ZrTe$_5$, Na$_3$Bi, GdPtBi, and  YbPtBi~~\cite{Huang2015, Xiong2015,  Hirschberger2016a,Zhang2016, Li2016, Armitage2018, Guo2018a}.
PHE was observed less often, in Cd$_3$As$_2$, VAl$_3$, Na$_3$Bi, as well as half-Heusler phases GdPtBi, YbPtBi, and DyPdBi~~\cite{Li2018e, Singha2018a, Liang2018b, Kumar2018,Guo2018a,Pavlosiuk2019}. To date, angular narrowing of negative LMR has been found only in Na$_3$Bi~~\cite{Xiong2015}.  

ScPtBi, isostructural with GdPtBi, YbPtBi and DyPdBi, but outstanding due to yhe very light and non-magnetic scandium atom, factors important for the topological non-triviality of electronic states.
It has been theoretically predicted to be a zero-gap semiconductor with inversion of electronic bands prerequisite for a $\mathbb{Z}_2$-type topological insulator~~\cite{Chadov2010a,Al-Sawai2010}. So far, only the observation of the weak antilocalization (WAL) effect and linear magnetoresistance suggested its topological non-triviality~~\cite{Hou2015b}. 
In this work, we synthesized the high-quality single crystals of ScPtBi and characterized them and their angle-dependent magnetotransport depending on the angle between current and applied magnetic field.
\section{Methods}
Single crystals of ScPtBi were grown by the self-flux method from high purity constituent elements (Sc 3N, Pt 5N, Bi 6N) taken in the atomic ratio: Sc:Pt:Bi as 1:1:15, put in an alumina crucible and sealed in a quartz ampule with argon pressure of 0.3\,bar. Such a prepared ampule was quickly heated up to 1150\,$^\circ$C at a rate of 50\,$^\circ$C/h and held at this temperature for 24 hours. Then it was slowly cooled down to 700\,$^\circ$C at a rate of 1\,$^\circ$C/h and the excess of Bi-flux was removed by centrifugation. Obtained single crystals were pyramid-shaped  and stable against air, and moisture. The largest of them had dimensions of 6$\times$3$\times$3\,mm$^3$ (see the right inset in Fig.~S1 in the Supplemental Material\cite{SupMat}).

Equiatomic chemical composition and phase homogeneity of crystals were confirmed by energy-dispersive X-ray analysis using a FEI scanning electron microscope equipped with an EDAX Genesis XM4 spectrometer (Supplemental Fig.~S1~~\cite{SupMat}). 
X-ray powder diffraction experiment was carried out on powdered single crystals using an X’pert Pro (PANanalytical) diffractometer with \mbox{Cu-K$\alpha$} radiation. Crystal structure refinement by Rietveld method was performed with the {\em Fullprof} package~~\cite{Rodriguez-Carvajal1993} (a diffractogram and results of the refinement are shown in Supplemental Fig.~S2A and Table~S1~~\cite{SupMat}). 
The results of analysis indicated that ScPtBi crystallizes within the $F\overline{4}3m$ space group with the lattice parameter $a = 6.4412\,$\AA, which differs very slightly from $a=6.50\,$\AA~ reported in Ref.~\onlinecite{Hou2015b}. 
Quality and orientation of single crystals were assessed by Laue diffraction, with a LAUE-COS (Proto) system (Laue pattern is shown in Supplemental Fig.~S2B~~\cite{SupMat}). 
Bar-shaped (0.19$\times$0.65$\times$1.73\,mm$^3$) specimen was cut from the oriented single crystal. Conventional four-probe method was used for electrical transport measurements in the temperature range from  0.05\,K to 300\,K and in the magnetic fields up to 14\,T, using a  PPMS platform (Quantum Design) equipped with dilution refrigerator. Electrical contacts were made from $50\,\mu$m-thick silver wires attached to the sample with silver paint. We took special care while preparing contacts, to avoid current jetting, which could cause spurious negative LMR, as described in Refs.~\onlinecite{Liang2018b,Pavlosiuk2019}. The same PPMS platform was used for specific heat measurements. 
\begin{figure*}[t]
	\includegraphics[width=0.9\textwidth]{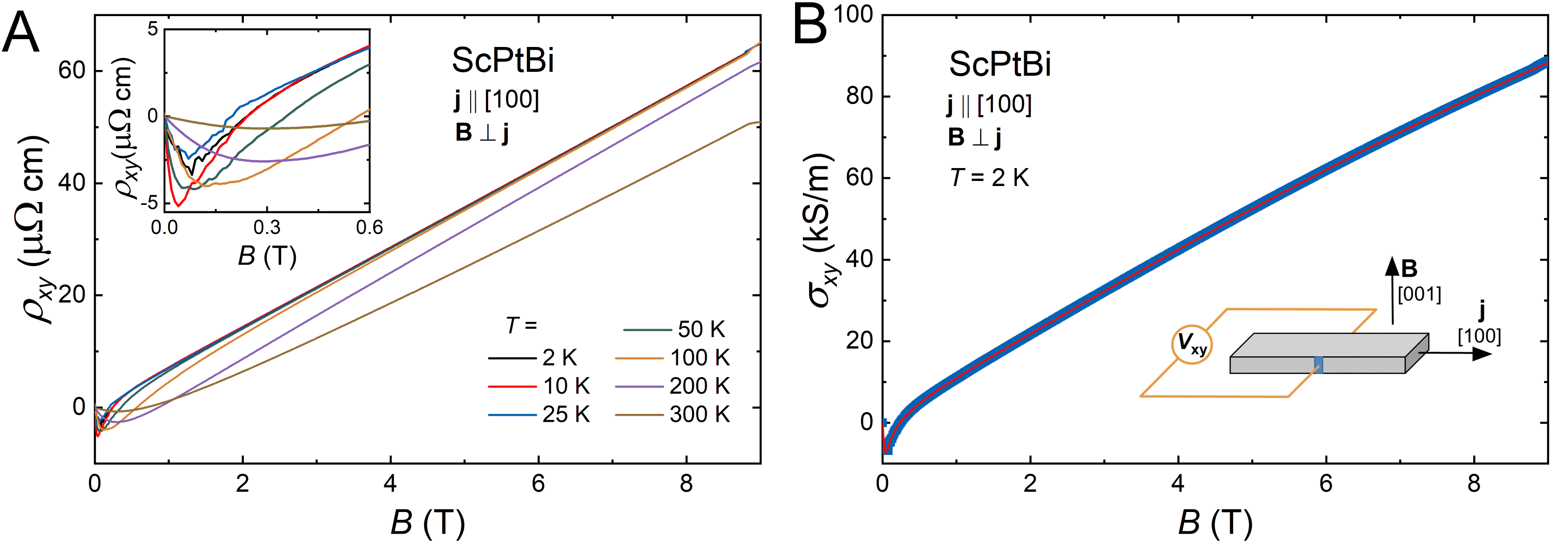}\vspace{-0.5cm}
	\caption{(A) Hall resistivity measured at several different temperatures as a function of magnetic field with ${\mathbf j} \parallel$[100] and ${\mathbf B} \parallel$[001]. The inset presents the Hall resistivity in weak magnetic field region. (B) Magnetic field dependent Hall conductivity measured at $T$= 2\,K. The red solid line represents the fit with Equation (1). Quality of the fit was excellent (with $R^2>$99.99). Inset shows the geometry of measurement.}
		\label{Hall2K_fig}
\end{figure*}

The electronic structure was calculated by the Full-Potential Local Orbital Minimum Basis (ver. 18) method~~\cite{Koepernik1999,Eschrig2003} 
 in the local density approximation, with the exchange correlation potential taken in the form of Perdew, Burke, and Ernzerhof (PBE)~~\cite{Eschrig2003,Perdew1996}. 
The band structure of the ordered ScPtBi was calculated in the full-relativistic mode in general gradient approximation, including the spin-orbit interaction.
For the disordered structures and for models with vacancies the band calculations were performed for the supercell of 24 atoms (Sc$_8$Pt$_8$Bi$_8$) in the full-relativistic and scalar-relativistic schemes, respectively. 
In the antisite models we interchanged two atoms from two different sublattices, per supercell. In the models with vacancies, one of atoms (Sc, Pt or Bi) was vacant per supercell.\vspace{-0.5cm}
\section{Results and discussion} \vspace{-4mm}
Contrary to prior theoretical predictions~~\cite{Chadov2010a,Al-Sawai2010}, our ScPtBi samples exhibited a metallic-like temperature dependence of electrical resistivity, $\rho_{xx}$, shown in Fig.~S4A of the Supplemental Material~~\cite{SupMat}. 
Moreover, zooming into the low-temperature region we observed below 0.7\,K an onset of a superconducting transition, with $\rho_{xx}$ finally decreasing to zero below 0.23\,K (see the inset to Fig.~S4A of the Supplemental Material~~\cite{SupMat}).  

The Hall resistivity isotherms, $\rho_{xy}(B)$, are nonlinear, especially in weak magnetic fields, where they have minima, indicating two or more different bands contributing to $\rho_{xy}$ (Fig.~\ref{Hall2K_fig}A and its inset; raw data before anti-symmetrization are shown in Supplemental Fig.~S5~~\cite{SupMat}). Previously reported rectilinear $\rho_{xy}(B)$ data for ScPtBi showed only one type of carriers contributing to the transport properties~~\cite{Hou2015b}, suggesting different electronic structures of our samples. Nonlinear $\rho_{xy}(B)$ have also been observed for several half-Heusler compounds: YbPtBi, LuPtBi, YPtBi, TbPdBi, and DyPdBi~~\cite{Mun2013,Hou2015,Pavlosiuk2016b,Xiao2018,Pavlosiuk2019}. Therefore, for a quantitative analysis of our data we used the three-band (the two-band model was tested as well, but it did not fit our data) Drude model of Hall conductivity:
\begin{equation}
\sigma_{xy}(B)=\sum_{i=1}^3\frac{e\,n_i \mu_i^2B}{1+(\mu_iB)^2},
\label{three-band-sigmaxy}
\end{equation}
summing up the conductivities of individual bands, with $n_i$ and $\mu_i$ denoting respectively the concentration and
mobility of carriers from the $i$th band ($e$ is the elementary charge).
We fitted Eq.~\ref{three-band-sigmaxy} to the experimental Hall conductivity data calculated as $\sigma_{xy}=\rho_{xy}/(\rho_{xx}^2+\rho_{xy}^2)$, appropriate because $\rho_{xy}\!\ll\!\rho_{xx}$.
The result of Eq.~\ref{three-band-sigmaxy} fitting to $\sigma_{xy}(B)$ obtained at $T=2$\,K is represented by red solid line in Fig.~\ref{Hall2K_fig}B. 
The same Eq.~\ref{three-band-sigmaxy} was fitted to the $\sigma_{xy}(B)$ data measured at other temperatures, the results are shown in Supplemental Fig.~S6~~\cite{SupMat}, and fit parameters are collected in Supplemental Table~S2~~\cite{SupMat}. 

Interestingly, charge mobilities derived for two of these bands are very high, especially that of an electron-like band: $\mu_2\approx2\times10^5$ $\rm{cm^2V^{-1}s^{-1}}$, which is the highest ever observed for pnictogen-bearing half-Heusler phases, and similar to those in topological semimetals ZrTe$_5$~~\cite{Chi2017a}, TaAs~~\cite{Zhang2017g} and WTe$_2$~~\cite{Ali2015}.  
\begin{figure*}
\includegraphics[width=\linewidth]{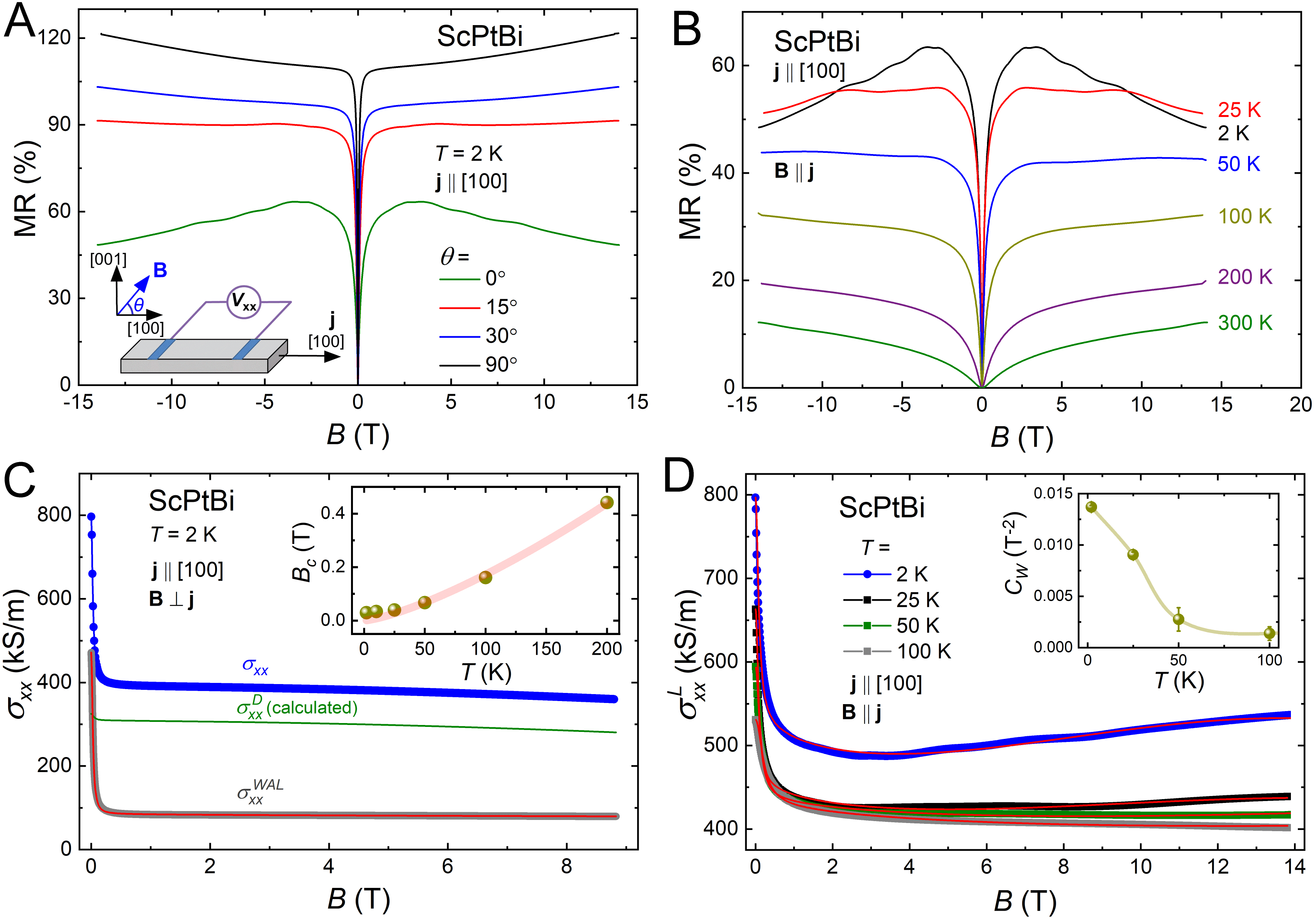}
	\caption{(A) MR($B$) isotherms measured at $T=2$\,K for different angles between current and magnetic field. Inset shows schematically the geometry of measurement. (B) LMR($B$) isotherms for several temperatures.
	(C)~Analysis of the transverse magnetoconductivity data obtained at $T=$\,2\,K. Blue points - experimental data, green solid line - $\sigma_{xx}^{D}(B)$ calculated with Eq.~(\ref{three-band-sigmaxx-model}). Grey solid curve represents the WAL contribution derived as described in the text, and red solid line curve shows the fit of Eq.~(\ref{WAL_tran}). Inset: temperature dependence of the critical field approximated by power law $B_c\!\propto\!T^{1.35}$. (D) Magnetic field dependence of the longitudinal magnetoconductivity for several temperatures. Red solid lines correspond to the fits of Eq.~(\ref{WAL_longitudinal}). Inset: temperature-dependent chiral coefficient (solid line is a guide for the eye).  
	\label{CMA}}
\end{figure*}

Looking for topological semimetal features in ScPtBi, we measured its $\rho_{xx}$ as a function of magnetic field, $B$, and calculated the magnetoresistance, MR $=(\rho_{xx}(B)/\rho_{xx}(0))\!-\!1$, for different angles, $\theta$, between $\mathbf B$ and electric current, $\mathbf j$ (as shown in  Fig.~\ref{CMA}A for $T=$\,2\,K). MR is the largest in transverse configuration (TMR), i.e. with $\mathbf B\,\perp\,\mathbf j$.   
When $\mathbf B$ is rotated ($\theta=90^{\circ}\rightarrow0^{\circ}$, as sketched in the inset to  Fig.~\ref{CMA}A), MR decreases monotonously, down to the lowest value in an LMR configuration ($\theta=0^{\circ}$). The field variation of LMR at 2\,K is fairly unusual: it increases abruptly with increasing $B$, reaches a maximum of 64\% in 2.75\,T, and decreases gradually in stronger fields. 

The magnetic field dependence of LMR changes prominently with temperature (Fig.~\ref{CMA}B). 
Negative slope of LMR($B$) at $T$\,=\,2\,K, gradually becomes positive as temperature increases. This behavior can be ascribed to the presence of at least two contributions:  increasing-in-field (dominating in weak fields) and decreasing-in-field (taking over in stronger fields). 

Finite LMR has been observed for many compounds, and when positive, it can be ascribed to anisotropic scattering~~\cite{Sondheimer1962} or anisotropy of the Fermi surface~~\cite{Pippard2009}. Also macroscopic inhomogeneities may lead to positive or negative LMR~~\cite{Hu2007}, but they are absent in our samples (Supplemental Fig.~S1~~\cite{SupMat}).  
LMR in topological semimetals is characterized by a distinct, additional negative contribution from CMA~~\cite{Son2013}. Therefore, we analyzed LMR in ScPtBi to check for its presence.   

MR measured for ScPtBi in both transverse and longitudinal configuration, demonstrates very sharp increase in weak magnetic field region (Fig.~\ref{CMA}A), that can be ascribed to WAL~~\cite{Hou2015b}. In stronger fields the increase of TMR becomes much slower. In order to describe this complex variation of TMR, we combined a WAL contribution with three-band Drude magnetoconductivity calculated as:
\begin{equation}
\sigma_{xx}^{D}(B)=\sum_{i=1}^3\frac{e\,n_i \mu_i}{1+(\mu_i B)^2},
\label{three-band-sigmaxx-model}
\end{equation}
using the concentrations and mobilities of charge carriers obtained in the analysis of Hall resistivity (cf. Supplemental Table S2~~\cite{SupMat}). $\sigma_{xx}^{D}(B)$ for 2\,K is drawn in  Fig.~\ref{CMA}C as a green solid line. Next, we subtracted $\sigma_{xx}^{D}(B)$ from the transverse electrical magnetoconductivity $\sigma_{xx}\;[=\rho_{xx}/(\rho_{xx}^2+\rho_{xy}^2)]$ (blue points). In this way we obtained the WAL contribution $\sigma_{xx}^{WAL}$ to TMR in ScPtBi, shown in  Fig.~\ref{CMA}C as gray solid line. Remarkably, we could very well approximate it with the equation proposed to describe WAL in Weyl semimetals:
\begin{equation}
\sigma_{xx}^{WAL}(B)=C_1^{qi}\frac{B^2\sqrt {B}}{B_c^2+B^2}+C_2^{qi}\frac{B_c^2B^2}{B_c^2+B^2}+\sigma_0,
\label{WAL_tran}
\end{equation}
where $\sigma_0$ is the electrical conductivity in zero magnetic field, $B_c$ stands for the critical field related to the phase coherence length, $l_{\phi}$, as $B_c\!\sim\!\hbar/(el_{\phi}^2)$~~\cite{Dai2016a}. Fitting this Equation to $\sigma_{xx}^{WAL}$ data (as shown by red line in Fig.~\ref{CMA}C), we obtained parameters listed in Supplemental Table S3A~~\cite{SupMat}.

According to Ref.\,\cite{Dai2016a} at low temperatures the low-field magnetoconductivity is proportional to $+\sqrt{B}$ in double-Weyl semimetals and to $-\sqrt{B}$ in single-Weyl semimetals. For $\sigma_{xx}$ of our sample at $T=$2\,K, the critical field, $B_c$,  is very small (0.029 T in Table S3), thus the first term of Eq.~\ref{WAL_tran} is very well approximated with $C_1^{qi}\sqrt{B}$. Therefore (since $C_1^{qi}<$0) the low-field $\sigma_{xx}$ is proportional to $-\sqrt{B}$, which allows us to conclude that ScPtBi is a single-Weyl semimetal. 

Analysis of WAL we performed for TMR at higher temperatures is shown in Supplemental Fig.~S7 and Table S3~~\cite{SupMat}. 
Temperature variation of critical field $B_c$, shown in the inset to  Fig.~\ref{CMA}C, can be well approximated with a power law $B_c=AT^p$, with $A=3.47\!\times\!10^{-4}$\,T/K and $p$\,=\,1.35. This value of $p$ is close to 1.5, theoretical value for electron-electron interaction, the principal mechanism of decoherence, indicating that our estimation of the WAL contribution is adequate.

In single-Weyl semimetals WAL is isotropic~~\cite{Lu2015}, contributing to both the longitudinal and transverse magnetoconductivity in the same way. Lorentz force acting on electrons in parallel magnetic and electric fields is zero, hence LMR is not affected by orbital magnetoresistance~\cite{Pippard2009}. 
Thus only contributions from WAL and CMA must be taken into account. 
We obtained the longitudinal magnetoconductivity $\sigma_{xx}^L(B)$  in ScPtBi as inverse of the electrical resistivity measured in longitudinal magnetic field: 
$\left.\sigma_{xx}^L(B)=1/\rho_{xx}(B)\right|_{\mathbf{B}\parallel\mathbf{j}}$  (plotted in  Fig.~\ref{CMA}D). Fitting these data with the formula appropriate for Weyl semimetals~~\cite{Kim2013a}:
\begin{equation}
\sigma_{xx}^L =\sigma_{xx}^{WAL}(B)\cdot(1+C_WB^2)+\sigma_n,
\label{WAL_longitudinal}
\end{equation}
(represented by red solid lines in  Fig.~\ref{CMA}D) where $\sigma_n=\sigma_{xx}^L(B=0)\!-\!\sigma_0$, gave the parameters gathered in Supplemental Table S3B~~\cite{SupMat}. 
This analysis reveals that the negative contribution to LMR that we observed may be due to CMA. It weakens with increasing temperature, as reflected  by the decrease of the chiral coefficient, $C_W$ (cf. the inset to  Fig.~\ref{CMA}D). 
The chiral coefficient that we derived (0.015\,T$^{-2}$ at 2\,K) is small, but very close to 0.021\,T$^{-2}$ observed in
isostructural YbPtBi, a well established topological semimetal~~\cite{Guo2018a}.

\begin{figure*}
\centering
	\includegraphics[width=0.85\linewidth]{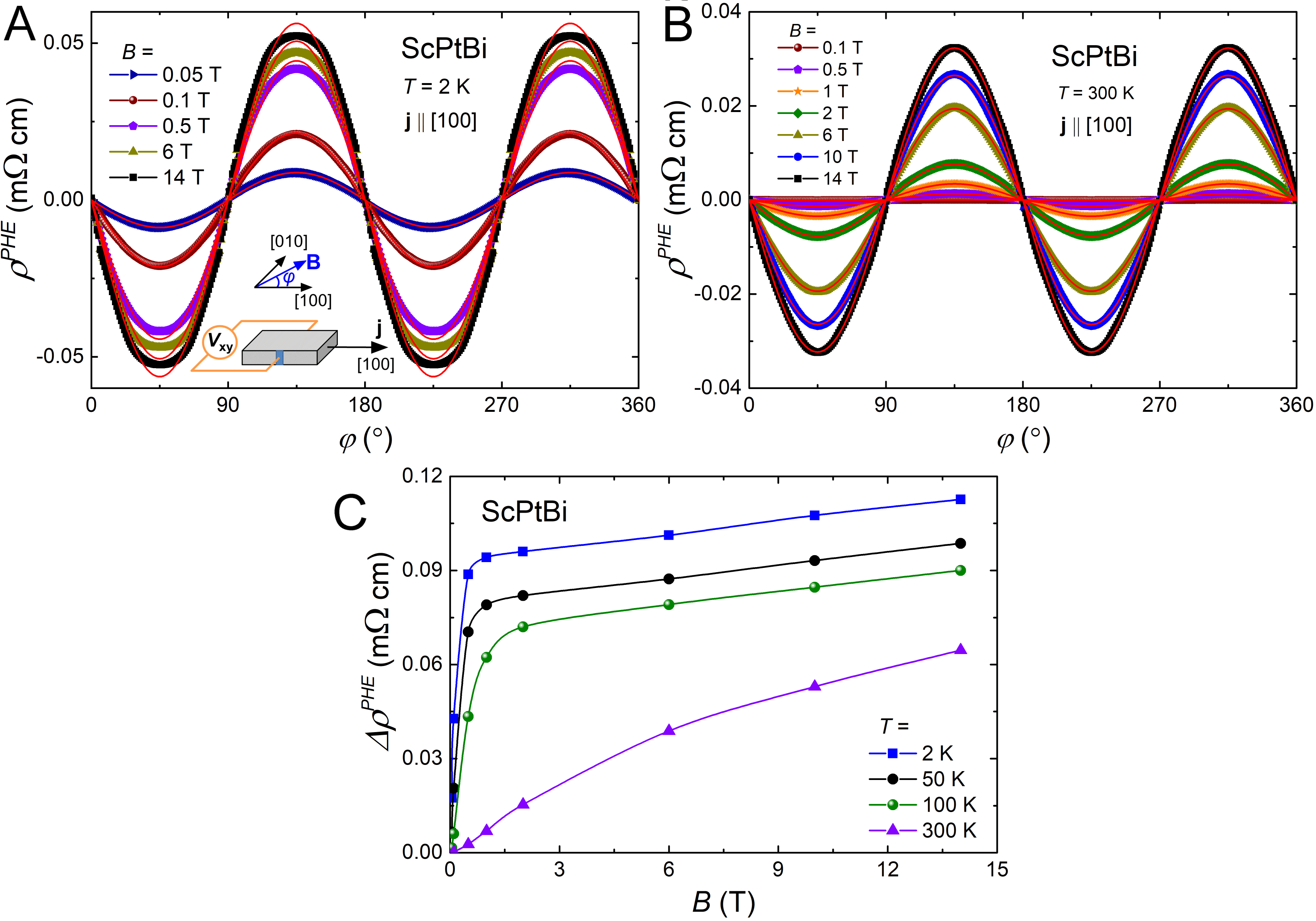}\vspace{-0.5cm}
	\caption{Angular dependence of the planar Hall resistivity measured in different external magnetic fields at 2\,K (A), and 300\,K (B), with experimental geometry sketched in inset to panel (A). Red solid lines represent the fits obtained using Eq.~(\ref{eq_PHE}). (C)~Magnetic field variations of the PHE magnitude observed at several different temperatures. Lines are guides for the eye. \label{PHE}}
\end{figure*}   
Next, we performed measurements of another indicator of CMA: the planar Hall resistivity, $\rho^{P\!H\!E}$, in the geometry shown in the inset to Fig.~\ref{PHE}A. Magnetic field was always in plane of the sample, voltage drop was recorded as a function of angle between magnetic field and current direction, $\phi$. $\rho^{P\!H\!E}$ depends on angle $\phi$ as follows~~\cite{Pippard2009,Burkov2017}:
\begin{equation}
\rho^{P\!H\!E}(\phi)=-\Delta\rho^{P\!H\!E}\sin2\phi,
\label{eq_PHE}
\end{equation}
where $\Delta\rho^{P\!H\!E}$ stands for the magnitude of PHE. We found that behavior of $\rho^{P\!H\!E}(\phi)$ in ScPtBi is typical, and persists in temperatures up to 300K (Fig.~\ref{PHE}B). 
\begin{figure*}[t]
\centering
	\includegraphics[width=0.77\linewidth]{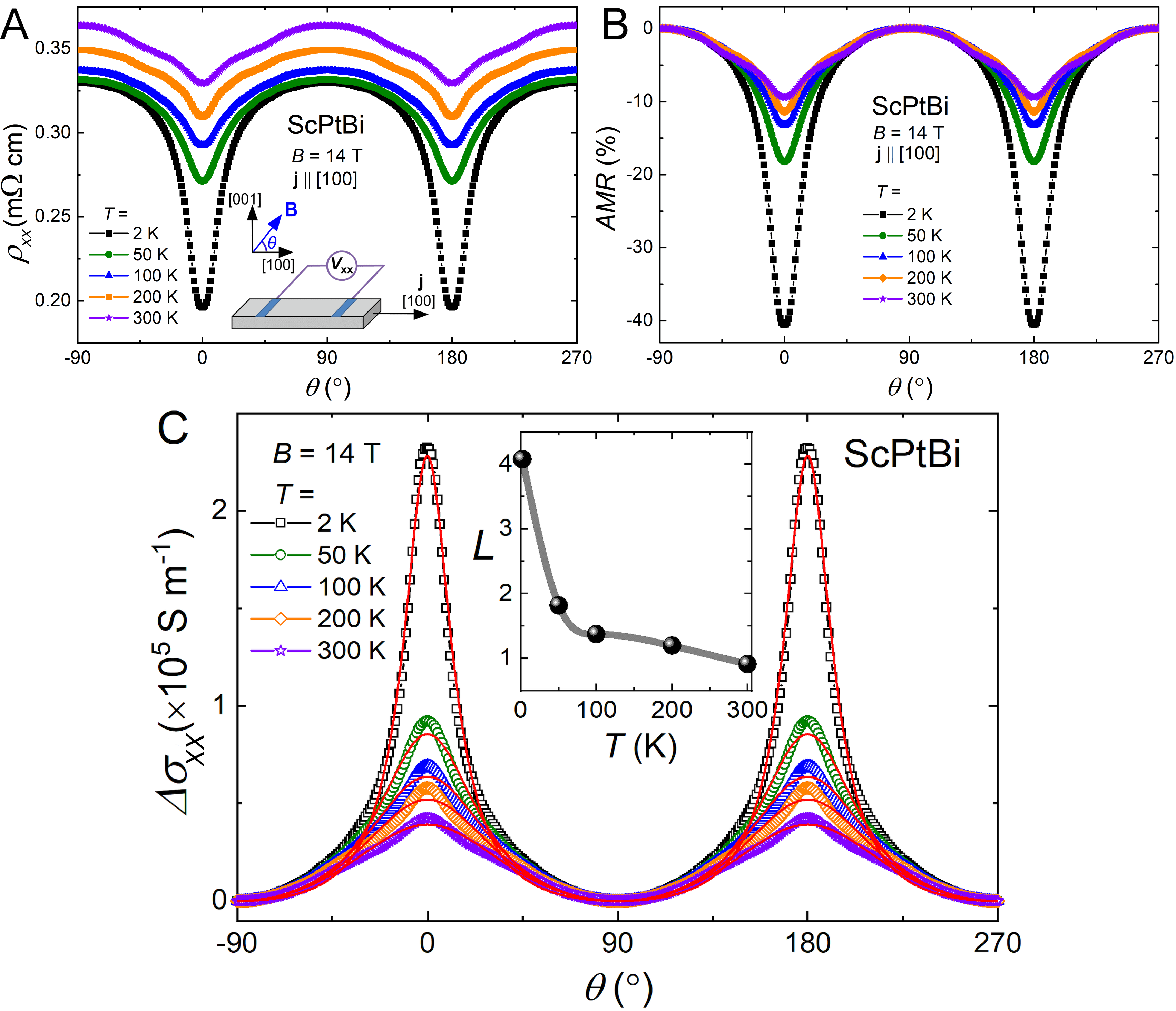}
	\caption{(A) Angular dependence of the electrical resistivity measured at several temperatures in a magnetic field of 14\,T. The inset shows the geometry of measurement. (B) Anisotropic magnetoresistance isotherms calculated from the data presented in (A). (C) Conductivity enhancement derived from the data shown in (A). Red lines represent the results of fitting Eq.~(\ref{ang_narrow_eq}) to the experimental data. The inset displays the temperature dependence of the narrowing parameter $L$ defined in the text. The solid line is a guide for the eye.    
	\label{narrowing}}
\end{figure*} 

Remarkably, PHE in ScPtBi is discernible even in very weak magnetic fields: $B=0.05$\,T at $T=2$\,K (see Fig.~\ref{PHE}A) and $B=0.1$\,T at $T=300$\,K ( Fig.~\ref{PHE}B), indicating this material as a possible directional magnetic field sensor. $\Delta\rho^{P\!H\!E}$ measured for ScPtBi increases with increasing the strength of applied magnetic field and decreases with increasing temperature, which is summarized in Fig.~\ref{PHE}C. 
The magnitude of PHE in ScPtBi is smaller than reported for other half-Heusler compounds~~\cite{Kumar2018,Pavlosiuk2019}, but larger than observed for VAl$_3$ or WTe$_2$~~\cite{Singha2018a,Li2019f}.  

The observation of PHE cannot be considered as sufficient proof of the existence of CMA, because any material which demonstrates anisotropic magnetoresistance (AMR) also shows PHE. AMR can have several different origins: ferromagnetism,  Fermi surface anisotropy~~\cite{Pippard2009}, and/or CMA~~\cite{Nielsen1983,Son2013}.  
In ferromagnets, AMR depends on $\theta$ as $\cos^2\theta$, while in materials with quasi-2D Fermi surface it depends on $\theta$ as $\left|\cos\theta\right|$~~\cite{Pippard2009}, where $\theta$ is  the angle between $\mathbf B$ and $\mathbf j$. 

In the presence of CMA, PHE must be accompanied by negative LMR and a particular angular dependence of AMR (also called angular narrowing of LMR)~~\cite{Burkov2017}. 
According to Burkov, the conductivity enhancement defined as $\Delta\sigma_{xx}\equiv\rho_{xx}^{-1}(\theta)-\rho^{-1}_{\perp}$ (where $\rho_{\perp}\!\equiv\!\rho_{xx}(90^{\circ})$), should depend on $\theta$ in the following way:
\vspace{-0.1cm}
\begin{equation}
\Delta\sigma_{xx}\propto\frac{L^2\cos^2\theta}{1+L^2\sin^2\theta},
\label{ang_narrow_eq}
\end{equation}
with $L=L_c/L_a$, where $L_c$ is the chiral charge diffusion length, and $L_a$ is the magnetic field related length scale~~\cite{Burkov2017}.

Thus, we measured $\rho_{xx}$ of ScPtBi as a function of $\theta$ in constant field $B=14$\,T at temperatures in the range from 2\,K to 300\,K (see  Fig.~\ref{narrowing}A). 
Next we calculated AMR\,$=\!(\rho_{xx}(\theta)-\rho_{\perp})/\rho_{\perp}$, which is plotted in  Fig.~\ref{narrowing}B, and obviously follows neither $\cos^2\theta$ nor $\left|\cos\theta\right|$. 

However the model proposed by Burkov~~\cite{Burkov2017} does not include WAL contribution to conductivity, which as we showed above is important in ScPtBi. Because in single-Weyl SMs $\sigma_{xx}^{WAL}$ is isotropic~~\cite{Lu2015}, it will cancel out in $\Delta\sigma_{xx}(\theta)$ defined just as a difference of conductivities: $\sigma_{xx}(\theta)-\sigma_{xx}(90^{\circ})$. Consequently, we calculated $\sigma_{xx}(\theta)=\rho_{xx}(\theta)/(\rho_{xx}^2(\theta)+\rho_{xy}^2(\theta))$, using  $\rho_{xx}(\theta)$ measured in field of 14\,T  (black symbols in Fig.~\ref{narrowing}A) and $\rho_{xy}(\theta)=\rho_{xy}(14\,{\rm T})\sin\theta$, where $\rho_{xy}(14\,{\rm T})$ were extrapolated linearly to $B=$14\,T, values of $\rho_{xy}(B)$ measured at each temperature (shown in Fig.~\ref{Hall2K_fig}A).

We fitted Eq.~\ref{ang_narrow_eq} to all $\Delta\sigma_{xx}(\theta)$ isotherms, as shown by red solid lines in  Fig.~\ref{narrowing}C. 
Very good fit obtained for the data collected at $T=2$\,K confirms the angular narrowing of LMR, and therefore the presence of CMA in ScPtBi. 

The fit parameter $L$ “determines the strength of the chiral anomaly induced magnetotransport effects”~~\cite{Burkov2017}.  It is the ratio of two length parameters:  $L_c=\sqrt{D\tau_c}$  and  $L_a=D/(\Gamma B)$, where $D$ is diffusion coefficient of carriers (increasing with $T$), $\Gamma$ - parameter inversely proportional to density of states at Fermi level (nearly $T$-independent), and $\tau_c$  – the chiral charge relaxation time, slowly (because of topological protection and conservation of chiral charge) decreasing with temperature. Therefore, $L=(L_c/L_a)\propto 1/\sqrt{D}$ should decrease with increasing temperature, as it is indeed shown in inset to Fig.~\ref{narrowing}C. 

With increasing temperature, quality of the fitting worsens, as the contribution of CMA  to LMR is reduced,  which is reflected in decrease of the fitted parameter $L$, $\Delta\rho^{P\!H\!E}(T)$ (Fig.~\ref{PHE}C) and $C_W(T)$ (inset to Fig.~\ref{CMA}D). 

Values and form of $\Delta\sigma_{xx}(\theta)$ we observed are very similar to those reported for the Weyl semimetal Na$_3$Bi~\cite{Xiong2015}, which further supports the presence of topologically non-trivial electronic structure in ScPtBi. 
\subsection*{Electronic structure calculations}
Earlier electronic structure  calculations have suggested that ScPtBi is a zero-gap semiconductor with inverted bands near the center of Brillouin zone, where the gap closes~~\cite{Chadov2010a,Al-Sawai2010}, similar to that of GdPtBi~~\cite{Hirschberger2016a}. 

\begin{figure*}
 \centering
	\includegraphics[width=0.95\linewidth]{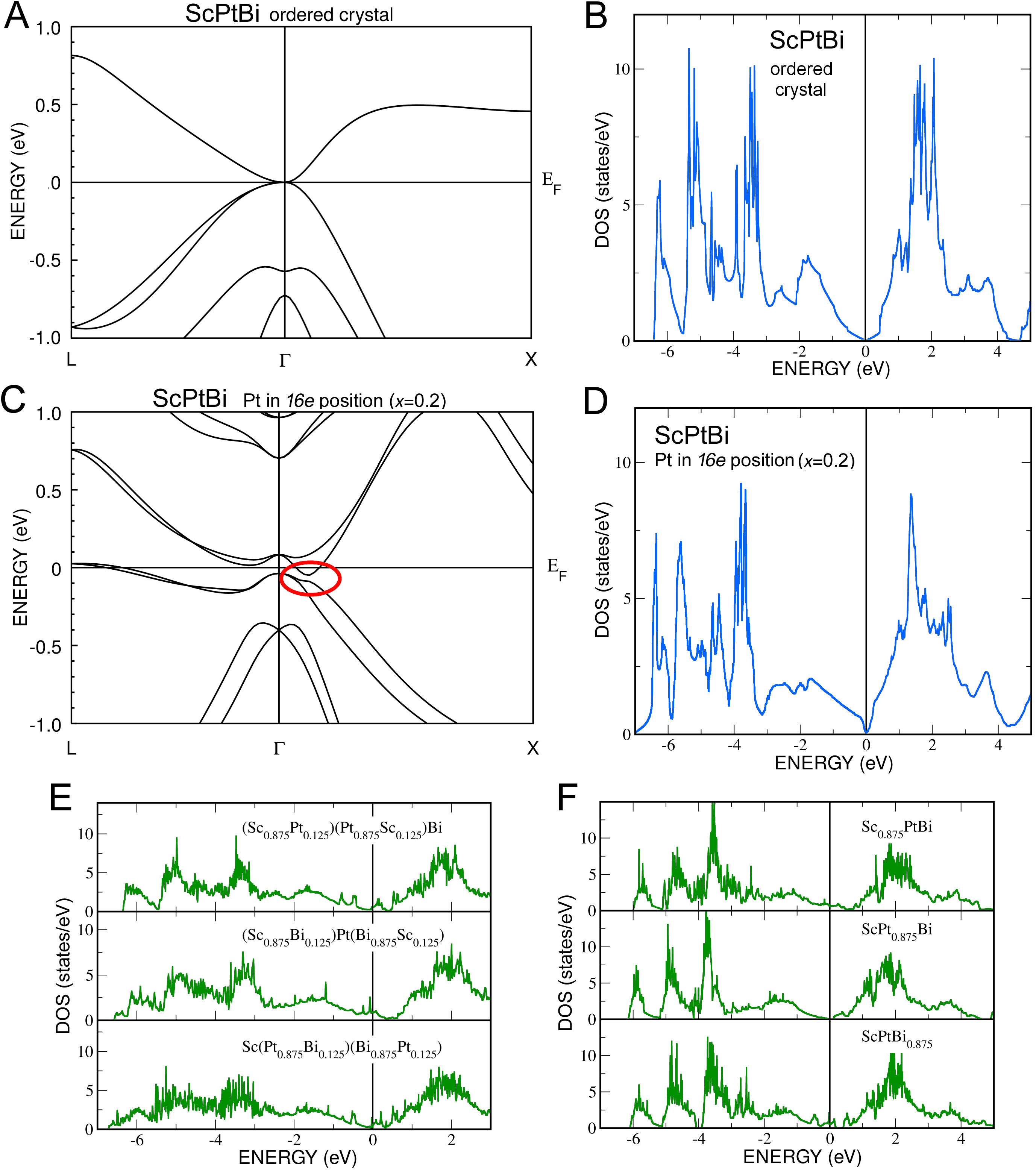}
	\caption{(A) Band~structure and (B) total DOS calculated (including spin-orbit interaction) for the ideal crystal structure of ScPtBi. (C) Band~structure and (D) total DOS calculated (including spin-orbit interaction) for the model with Pt occupying the $16e$ Wyckoff position. The red ellipse in C marks a gap between inverted bands where Dirac points may occur. (E,F) Total DOS for models: (E) with different antisites, and (F) with different vacancies. 
	\label{ele_stru}}
\end{figure*}

We got very similar results for a model with perfectly ordered ScPtBi crystal structure. The obtained band structure is shown in  Fig.~\ref{ele_stru}A and the density of states (DOS) as a function of energy has zero value at the Fermi level,  as shown in Fig.~\ref{ele_stru}B. This structure is, however, in disagreement with our experimental data (metallic-like resistivity - Supplemental Fig.~S4A, and specific-heat analysis in Supplemental Material~~\cite{SupMat}), which directly point to a substantial DOS at the Fermi level ($E_{\rm F}$). 

There has recently been proposed a model of half-Heusler crystal structure with Pt atom in Wyckoff position 
4$c$ ($^1\!/\!_4,^1\!/\!_4,^1\!/\!_4$) split in randomly quarter-occupied 16$e$ position ($x, x, x$)~~\cite{Synoradzki2019}. 
Refinement of X-ray diffraction pattern revealed that this model describes structure of our ScPtBi crystals equally well as the model of ideal half-Heusler (cf. Supplemental Table~S1~~\cite{SupMat}). 
 We incorporated this model in our electronic structure calculations and found that relatively big change of Pt position (to $x$ = 0.2, i.e. 0.05 shift from 4$c$) results in finite DOS($E_{\rm F}$) equal to 0.069 st./(eV f.u.) (see Figs.~\ref{ele_stru}C and \ref{ele_stru}D). Our ScPtBi crystal structure refinement (Supplemental Table S1) using that model gave $x$ = 0.242 for 16$e$ position of Pt atoms, i.e. only 0.008 deviation from $x={^1\!/\!_4}$ of 4$c$ position, much smaller than needed to sufficiently affect DOS. Therefore, this model alone does not suffice to explain semimetallic nature of ScPtBi. Interestingly, band structure for this model contains three bands crossing Fermi level, and an gap (of $\approx50$\,meV) between inverted bands, where Dirac points may occur (marked with a red ellipse in Fig.~\ref{ele_stru}C), indispensable for realization of Weyl semimetal. 
 
GdPtBi is a Weyl semimetal with Weyl nodes induced by external magnetic field. Magnetic field is not taken into account in {\em ab initio} band structure calculations (showing zero-gap semiconductor), and in zero-field GdPtBi does not show metallic $\rho_{xx}(T)$~~\cite{Hirschberger2016a}. Therefore, it is the influence of magnetic field on electronic structure, acting as a source of finite DOS$(E_{\rm F})$, that is a prerequisite to semimetallic state~~\cite{Hirschberger2016a}. For ScPtBi metallic-like $\rho_{xx}(T)$ is observed even in zero magnetic field, which allows us to discard that scenario.

The ideal half-Heusler structure consists of three sub-lattices, each fully occupied by atoms of only one kind. But in real, even apparently perfect single crystals, different defects, such as antisites or vacancies, always occur. Thus, we examined if structural defects can be alternative sources of finite DOS$(E_{\rm F})$ in ScPtBi. 

Calculations for models of ScPtBi with different antisites indeed brought finite DOS$(E_{\rm F})$ (see  Fig.~\ref{ele_stru}C) and thus profound changes in the electronic structure. 
For one Sc-Bi antisite per supercell, DOS$(E_{\rm F})$ is 0.83\,st./(eV\,f.u.), for a Sc-Pt antisite: 0.60\,st./(eV\,f.u.) and for Pt-Bi: 0.58\,st./(eV\,f.u.), where f.u. stands for formula-unit. 
Vacancies also lead to non-zero values of DOS$(E_{\rm F})$. One Pt- or Bi-vacancy per supercell yielded the DOS shown in  Fig.~\ref{ele_stru}F, which at the Fermi level has values of 0.67\,st./(eV\,f.u.), and 0.57\,st./(eV\,f.u.), respectively (but zero for a Sc-vacancy). 

All these finite DOS$(E_{\rm F})$ values are significantly larger than 0.09\,st./(eV\,f.u.) obtained from the specific heat results (shown in Supplemental Material~~\cite{SupMat}), most likely due to smaller degree of disorder in our real crystal than that assumed in models tested. 
However, regardless of antisite type and for Pt- or Bi-vacancies, as well as for the structure with Pt occupying $16e$ Wyckoff position, we always obtained semimetallic electronic structure. Additionally, in all these cases, Fermi level crosses both hole- and electron-like bands (Fig.~\ref{ele_stru}C and Supplemental Fig.~S8~~\cite{SupMat}), which is in agreement with the results of our Hall resistivity measurements (Fig.~\ref{Hall2K_fig} and Table~S2~~\cite{SupMat}).
Thus, both antisites and vacancies may be important sources of non-vanishing DOS at Fermi level in ScPtBi, and very likely also in other half-Heusler semimetals. Our calculations for antisite or vacancies did not reveal features of topological semimetal close to $E_{\rm F}$, but those for the structure with Pt at $16e$ position revealed a gap between inverted bands where Dirac points might occur. ScPtBi may become Weyl semimetal upon application of magnetic field, splitting such Dirac point in two Weyl nodes, alike in Na$_3$Bi or GdPtBi.  Xiong et al. proposed Zeeman interaction to be responsible for such a splitting in Na$_3$Bi~~\cite{Xiong2015}, whereas for GdPtBi, both the Zeeman effect~~\cite{Hirschberger2016a} and, alternatively, the exchange field due to the magnetic moments of Gd~~\cite{Shekhar2018}, have been suggested as direct causes of formation of Weyl nodes. In the case of ScPtBi, despite being isostructural with GdPtBi, the absence of atoms with strong magnetic moments leads to exclusion of the exchange field, and points to Zeeman splitting as the origin of Weyl nodes. 

\section{Conclusion} Our examination of magnetotransport in ScPtBi revealed non-linear Hall resistivity and large, positive and non-saturating TMR. Fitting the three-band Drude model to those data indicated ultrahigh carrier mobility in one of the bands. The low-field behavior of TMR could be well modeled with a WAL contribution.  

Importantly, upon rotation of the sample in magnetic field, passing from a transverse to a longitudinal configuration, the slope of MR($B$) changes from positive to negative (in fields above 3\,T). This reflects the significant negative component in LMR, a basic hallmark of CMA. Moreover, we observed in our sample two other crucial signs of CMA, namely the planar Hall effect and angular narrowing of negative LMR. With increasing temperature, particular parameters of all these properties: the chiral coefficient in longitudinal magnetoconductance, the magnitude of PHE and the chiral charge diffusion length clearly diminished, reflecting weakening of the CMA effect. Our analysis indicates that ScPtBi is a single-Weyl semimetal. 

Our unique observation of all three hallmarks of CMA simultaneously in one material, allows us to conclude that ScPtBi is the new representative of topological semimetals. 

The superconductivity we observed in ScPtBi makes this system even more interesting and worth further studies, due to conjunction with topologically non-trivial states, which may be important for quantum computing. \\\\
\noindent {\bf Acknowledgement} \\ Work was supported by the National Science Centre of Poland (2015/18/A/ST3/00057 to D.K. and P.W.), and the Foundation for Polish Science (START 66.2020 to O.P.). \clearpage
%

\clearpage
\setcounter{figure}{0}\setcounter{page}{0}\setcounter{equation}{0}
\renewcommand{\thefigure}{S\arabic{figure}}
\renewcommand{\thepage}{S-\arabic{page}}
\renewcommand{\theequation}{S\arabic{equation}}
\noindent\begin{large}Supplemental Material for\end{large}\\\\
\begin{large}{\bf Magnetotransport signatures of chiral magnetic anomaly in the half-Heusler phase ScPtBi}\\\\ 	
Orest Pavlosiuk$^1$, Andrzej Jezierski$^2$, Dariusz Kaczorowski$^1$ and Piotr Wi\'sniewski$^1$
\end{large}\\\\
$^1${\it Institute of Low Temperature and Structure Research,
	Polish Academy of Sciences,\\ 50-422 Wroc{\l}aw, Poland}\\
$^2${\it Institute of Molecular Physics, Polish Academy of Sciences, 60-179 Poznań, Poland}\\
\subsection*{Contents} 
\noindent Figure S1 - Energy-dispersive X-ray spectrum.  \\
Powder and Laue X-ray diffraction results (incl. Figure S2 and Table S1). \\
Specific heat capacity results (incl. Figure S3). \\
Electrical resistivity and transverse magnetoresistance results (incl. Figure S4).\\
Figure S5 - Raw data of Hall resistance. \\
Three-band model analysis of Hall conductivity (incl. Figure S6 and Table S2). \\
Weak antilocalization analysis and CMA parameters for different temperatures\\
(incl. Figure S7 and Table S3)\\
Supplementary results of electronic structure calculations (incl. Figs. S8 and S9). \\
Supplementary references.
\clearpage
\begin{figure*}[h]
	\includegraphics[width=0.7\textwidth]{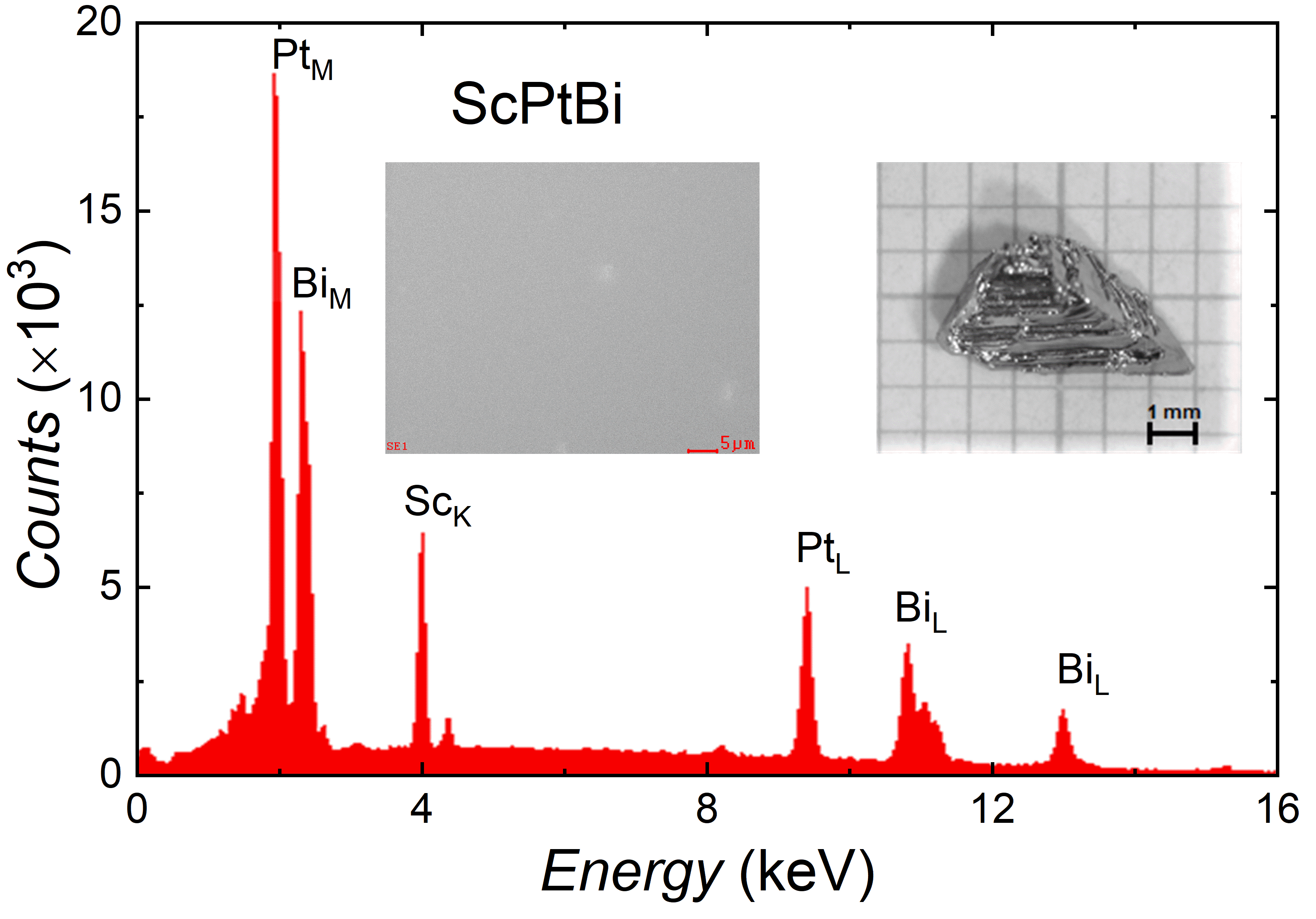}
	\caption{ Energy-dispersive X-ray spectrum (EDS), SEM image (left inset) and photograph (right inset) of ScPtBi single crystal. Chemical composition determined from the spectrum is close to the equiatomic: Sc$_{33.33}$Pt$_{34.03}$Bi$_{32.64}$.}\label{ESD_img}
\end{figure*}\vspace{-1cm}
\subsection*{Powder and Laue X-ray diffraction results}
\begin{figure*}[h]
	\includegraphics[width=0.55\textwidth]{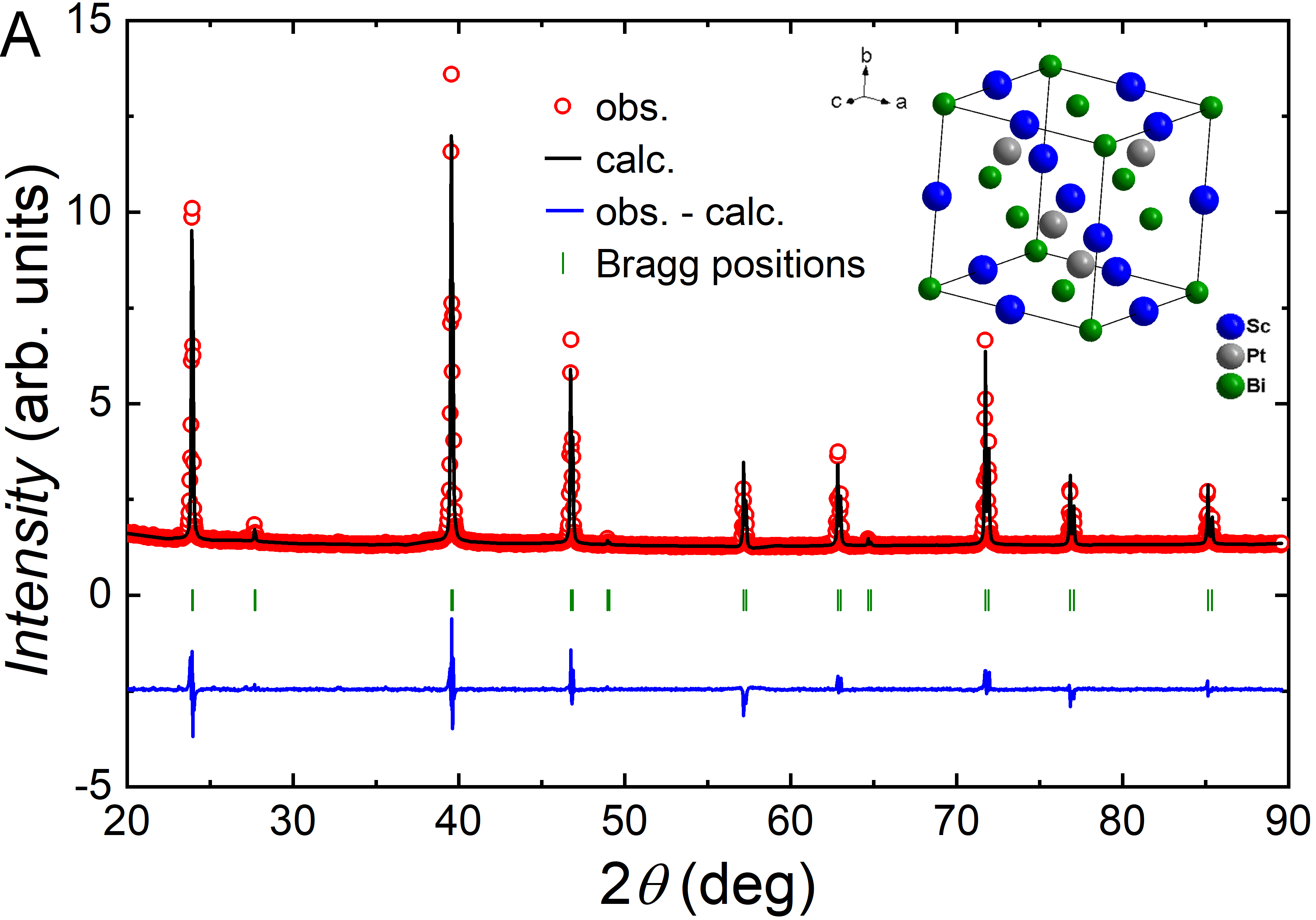}
	\includegraphics[width=0.43\textwidth]{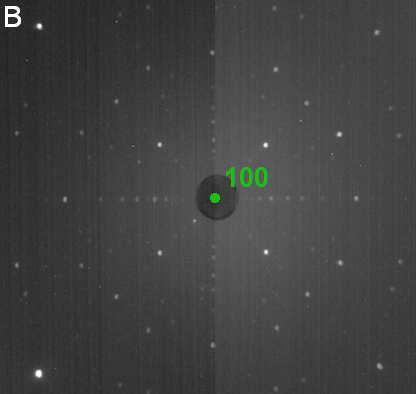}
	\caption{(A) X-ray diffractogram of powdered single crystals of ScPtBi. Inset shows the unit cell. (B) Laue diffraction pattern for ScPtBi single crystal taken with incident X-ray beam along the [100] crystallographic direction.}
		\label{Xray_img}
\end{figure*}

\vspace{-0.5cm}
Figure~\ref{Xray_img}A demonstrates an X-ray diffractogram of powdered ScPtBi (collected at 300\,K) and results of Rietveld refinement of its crystal structure. All obtained Bragg peaks could be indexed with the $F\overline{4}3m$ space group, typical for half-Heusler phases. Crystal structure data obtained from the refinement are gathered in Table S1, where {\em model 1} represents ideal half-Heusler crystal structure.\\
Several models of crystal structure have been taken into consideration in the refinement. Following the idea from [1], we replaced fully occupied position of Pt atom 4$c$ ($^1\!/\!_4,^1\!/\!_4,^1\!/\!_4$) with quarter-filled position 16$e$ ($x, x ,x$), this is denoted as {\em model 2} in Table S1. Our attempts to use other models which consider different type of crystal structure disorder (vacancies and antisites) did not yield statistically significant results, most likely because of accuracy limitations of our standard X-ray diffractometer.\\
Laue backscattering technique confirmed high quality of studied single crystal and allowed precise orientation and cutting of prism-shaped samples for electronic-transport measurements. Figure~\ref{Xray_img}B shows an exemplary Laue pattern.
\begin{table*}[b]
\raggedright{\textbf{TABLE S1} Crystal data and structure refinement results for ScPtBi.}\\\vspace{4mm} 
\begin{ruledtabular}
	\begin{tabular}{lll}
&{\em model 1}& {\em model 2}\\\colrule
space group	& \multicolumn{2}{c}{$F\overline{4}3m$}\\
lattice parameter, A&	6.441268(36)&	6.441257(37)\\
formula units per unit cell	&\multicolumn{2}{c}{4}\\
Sc Wyckoff position&\multicolumn{2}{c}{4$b$ (1,1,1) (fully occupied)}\\
Pt Wyckoff position&	4$c$ ($^1\!/\!_4,^1\!/\!_4,^1\!/\!_4$) & 16$e$ ($x, x, x$)\\
 &   (fully occupied)& (occupied in 25 {\%})\\
 &&     with $x$=0.2421(25)\\
Bi Wyckoff position &\multicolumn{2}{c}{4$a$ (0,0,0) (fully occupied)}\\
$B_{\rm iso}$ Sc	&1.65(38)	&1.53(38)\\
$B_{\rm iso}$ Pt	&0.82(20)&	0.63(20)\\
$B_{\rm iso}$ Bi	&0.09(17)	&0.08(17)\\
Bragg $R$-factor, \%	&5.33	&5.22\\
$R_f$-factor, \%	&3.21	&3.24\\
	\end{tabular}
\end{ruledtabular}
	\label{Xray_table}
\end{table*} 
\clearpage
\subsection*{Specific heat capacity}\vspace{-0.5cm}
\begin{figure*}[h] 
	\includegraphics[width=0.55\textwidth]{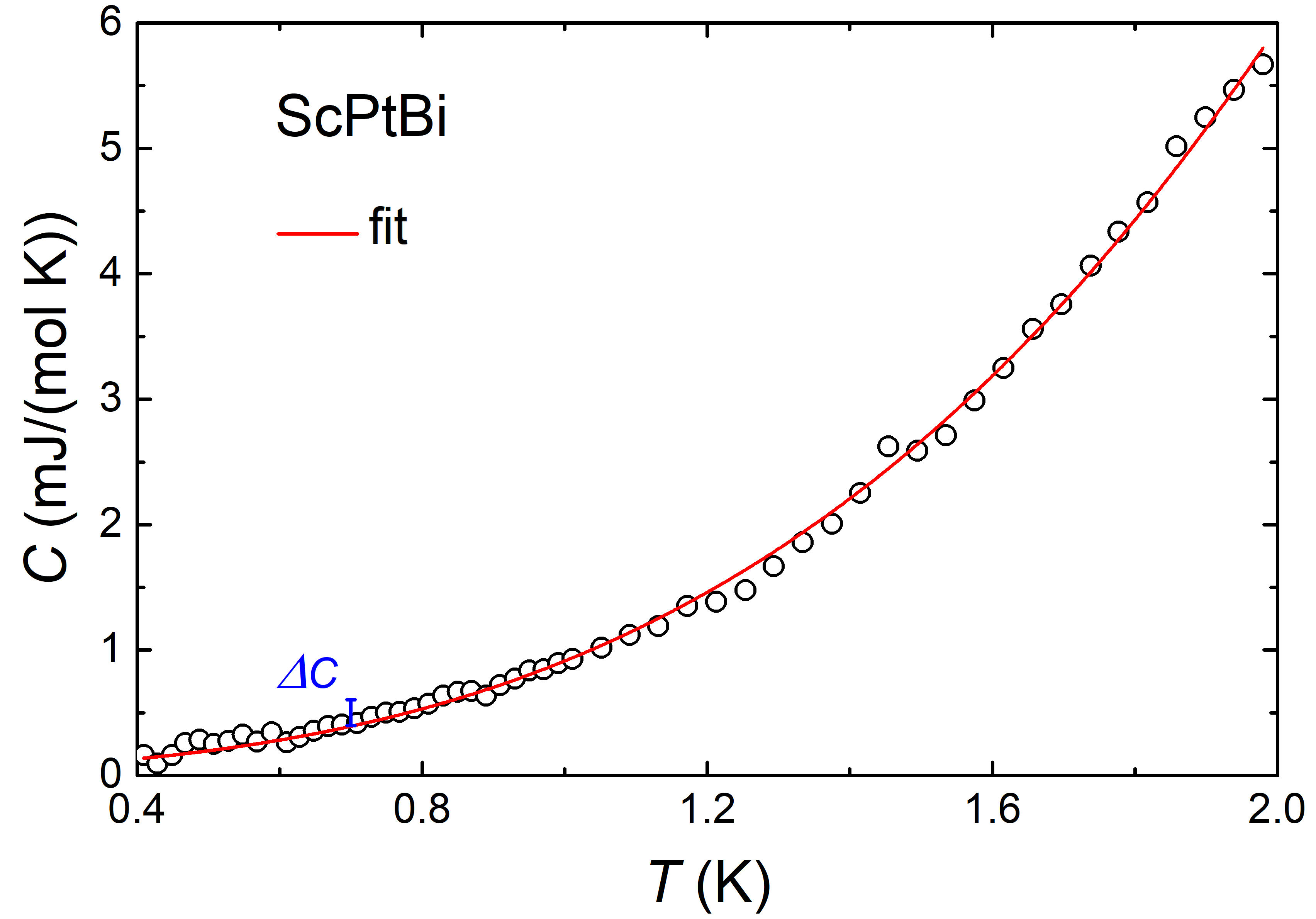}\vspace{-0.5cm}
	\caption{Temperature dependent specific heat of ScPtBi. 
	Red solid line corresponds to the fit with  $C=\gamma T+\beta T^3$ equation. 
	Blue bar corresponds to the specific heat jump expected for a conventional superconductor.}
		\label{HC_img}
\end{figure*}
Temperature dependence of specific heat (Fig.~\ref{HC_img}) shows no clear anomaly, which could be attributed to the superconducting phase transition. This is common feature for all half-Heusler superconductors which contain rare earth elements and bismuth  [2–6]. $C(T)$ is well described by a standard expression $C=\gamma T+\beta T^3$ , where $\gamma T$  corresponds to electronic contribution and $\beta T^3$ to phonon contribution to the total specific heat. The obtained fitting parameters are: $\gamma$= 0.22\,mJ/(mol\,K$^2$) and $\beta$= 0.69\,mJ/(mol\,K$^4$). From the value of $\beta$ we obtained Debye temperature: $\Theta_{\rm D}= (36 R\pi^4/5\beta)^{1/3}$= 204\,K, which is similar to those reported for other half-Heusler phases  [3,4]. 
Sommerfeld coefficient, $\gamma$, is proportional to density of states (DOS) at Fermi level: DOS$(E_{\rm F})=3\gamma(\pi k_{\rm B})^{-2}$, where $k_{\rm B}$ is Boltzmann constant. Using this formula we obtained DOS$(E_{\rm F})$= 0.09 st./(eV f.u), which is equivalent to 1.41$\times 10^{47}$\,st./(eV m$^{-3}$). In turn, carrier concentration, $n$, is proportional to DOS$(E_{\rm F})$ $(n =\, ^2\!/\!_3\,{\rm DOS}(E_{\rm F})$. 
Assuming that Fermi energy of ScPtBi is similar to that of YPtBi (estimated between 13 and 71 meV  [6,7]), we got $n \approx 10^{19}$\,cm$^{-3}$. Its order of magnitude is in very good agreement with carrier concentration values obtained from the three-bands model analysis of the Hall conductivity (Table S2). 
Considering that mechanism of superconductivity in ScPtBi can be described by BCS theory, we calculated the value of jump in $C(T_{\rm c})$, which equals to $\Delta C=1.43\,\gamma T_{\rm c}$= 0.21\,mJ/(mol\,K). This jump, shown as a blue error bar in Fig.~\ref{HC_img}, is very small, and thus could not be resolved in our experiment.
\clearpage
\subsection*{Electrical resistivity and transverse magnetoresistance}
\begin{figure*}[h]
	\includegraphics[width=0.49\textwidth]{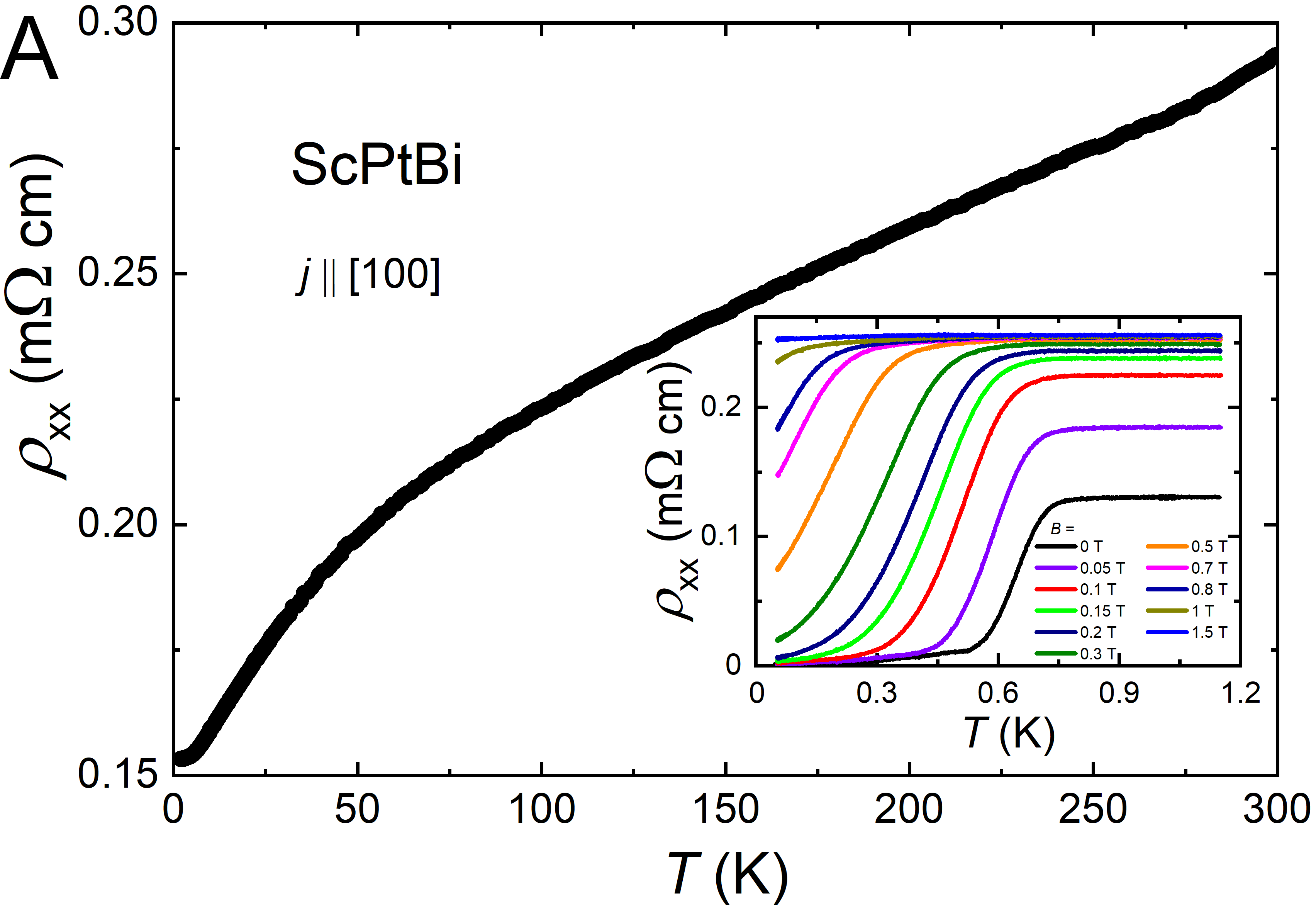}
	\includegraphics[width=0.49\textwidth]{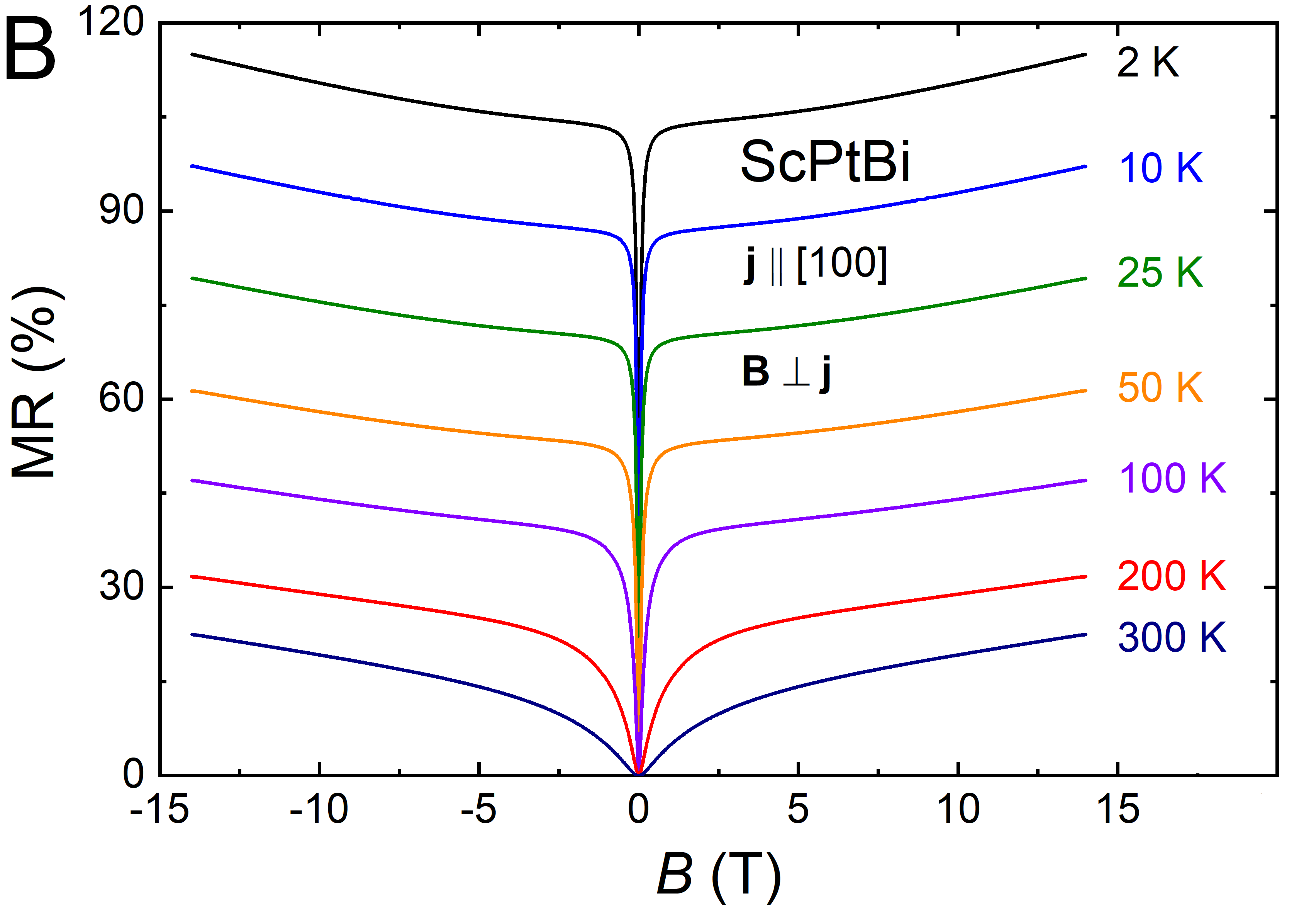}\vspace{-0.5cm}
	\caption{(A) Temperature dependence of the electrical resistivity measured with electric current ${\mathbf j} \parallel$ [100]. Inset shows electrical resistivity in low temperature range measured in magnetic fields of different strength aapplied transverse to the current. (B) Transverse MR measured at several different temperatures as a function of magnetic field.}\label{S4_fig}
\end{figure*}
\noindent Electrical resistivity, $\rho_{xx}$, of ScPtBi exhibits metallic-like temperature dependence (Fig.~\ref{S4_fig}A). Shape of the $\rho_{xx}(T)$ curve and the ratio $\rho_{xx}$(300\,K)/$\rho_{xx}$(2\,K)\,$\approx$2 are similar to those reported before  [8], however distinctly different from $\rho_{xx}(T)$ of GdPtBi, which electronic structure [9] was  supposed to closely resemble that of ScPtBi. Moreover, below 0.7\,K, $\rho_{xx}$ decreases sharply by about 90\%, then, below 0.5\,K, it undergoes another less abrupt decrease, down to zero value at $T=$0.23\,K. Similar two-step superconducting transitions have previously been reported for numerous half-Heusler superconductors  [2].
The critical temperature, $T_{\rm c}$, decreases with increasing strength of applied magnetic field, and at 0.05\,K the superconductivity becomes totally suppressed by a magnetic field of 1.5\,T (inset to Fig.~\ref{S4_fig}A). We thoroughly surveyed the literature and found that no binary compound of Sc, Pt and Bi is superconducting with $T_{\rm c}$ similar to that observed for our ScPtBi sample. Moreover, scanning electron microscopy with EDS analysis did not reveal any non-homogeneities in our crystals. Therefore, we can safely exclude the superconductivity originating from an impurity phase. 

Slightly different residual resistivities (0.16\,m$\Omega$\,cm in panel (A) and 0.13\,m$\Omega$\,cm in the inset) occur because these data were obtained in two resistance measurements. The first  was done with PPMS in its basic configuration (providing $T\ge$1.8 K), in the second one we used PPMS with dilution refrigerator, and sample  with fresh electrical contacts mounted on another sample holder. Distances between voltage contacts were slightly different in both cases, and the limited accuracy of their measurement caused this resistivity discrepancy. 

Transverse magnetoresistance of ScPtBi (TMR, ${\mathbf j} \parallel$[100], ${\mathbf B} \parallel$[001]), as a function of magnetic field at several temperatures from 2\,K to 300\,K, is shown in Fig.~\ref{S4_fig}B. It is positive in the entire covered temperature range and does not saturate in strong magnetic fields. It increases very abruptly in weak magnetic fields: for example, at $T$= 2\,K it reaches 100\% already in $B$= 0.35\,T. Values of $\rho_{xx}$ underlying these TMR curves were used for WAL analyses of magnetoconductivity for $T$= 2\,K (viz. Fig.~2C), and for higher temperatures (Fig.~S6 and Table S3A). 
%
\subsection*{Raw data of Hall resistance}
\begin{figure*}[h]
	\includegraphics[width=0.9\textwidth]{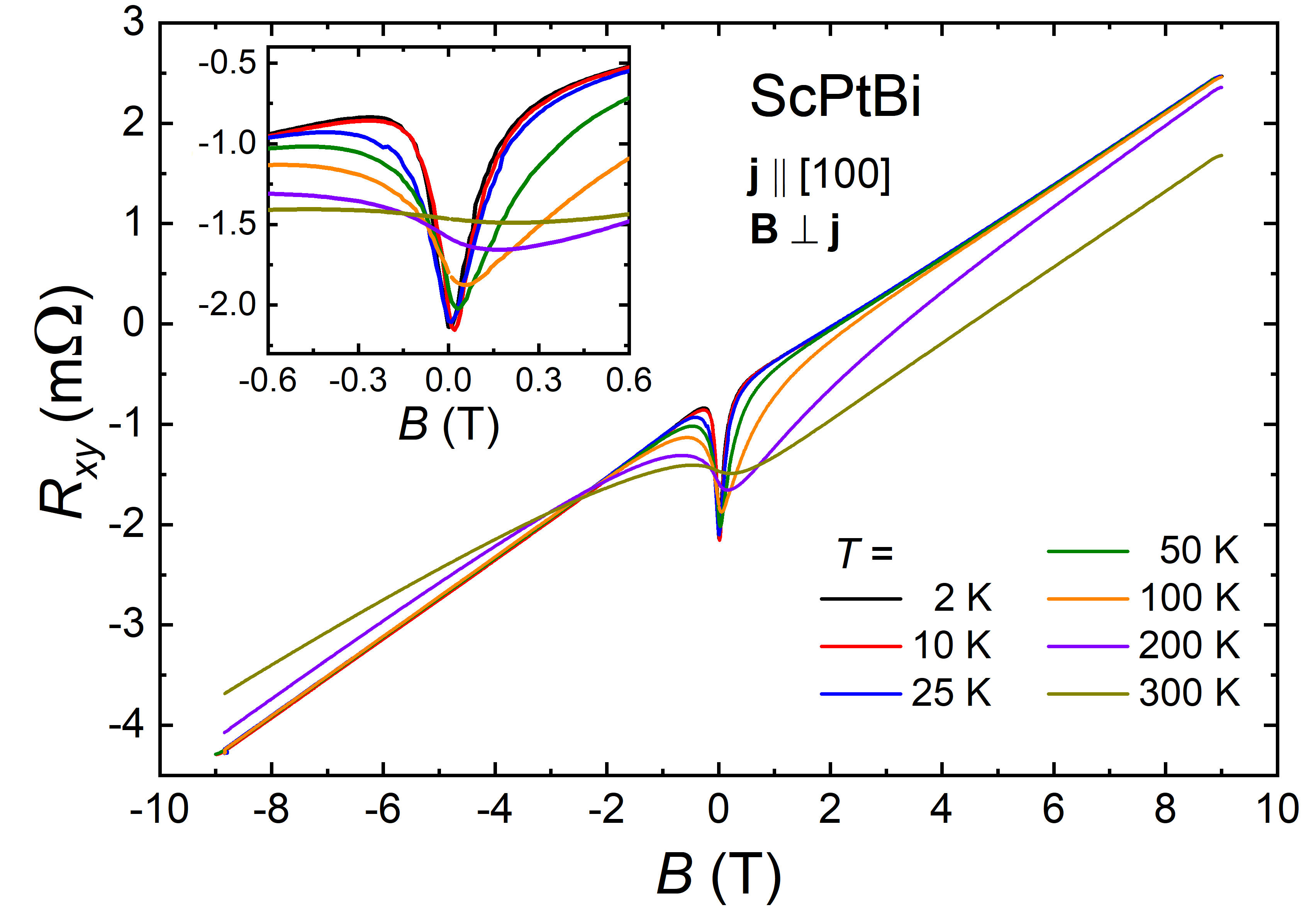}\vspace{-0.5cm}
	\caption{Hall resistance measured at several temperatures with electric current ${\mathbf j} \parallel$ [100]. Inset shows zoom in weak magnetic field range. Hall resistivity data shown in Fig.~1A of the main text were obtained through anti-symmetrization of data shown here, as:\\ $\rho_{xy}(B)=t(R_{xy}(B)-R_{xy}(-B))/2$, where $t$ was the thickness of the sample.}\label{Raw_Hall_fig}
\end{figure*} 
\subsection*{Three-band model analysis of Hall conductivity}
\begin{figure*}[h]
	\includegraphics[width=0.32\textwidth]{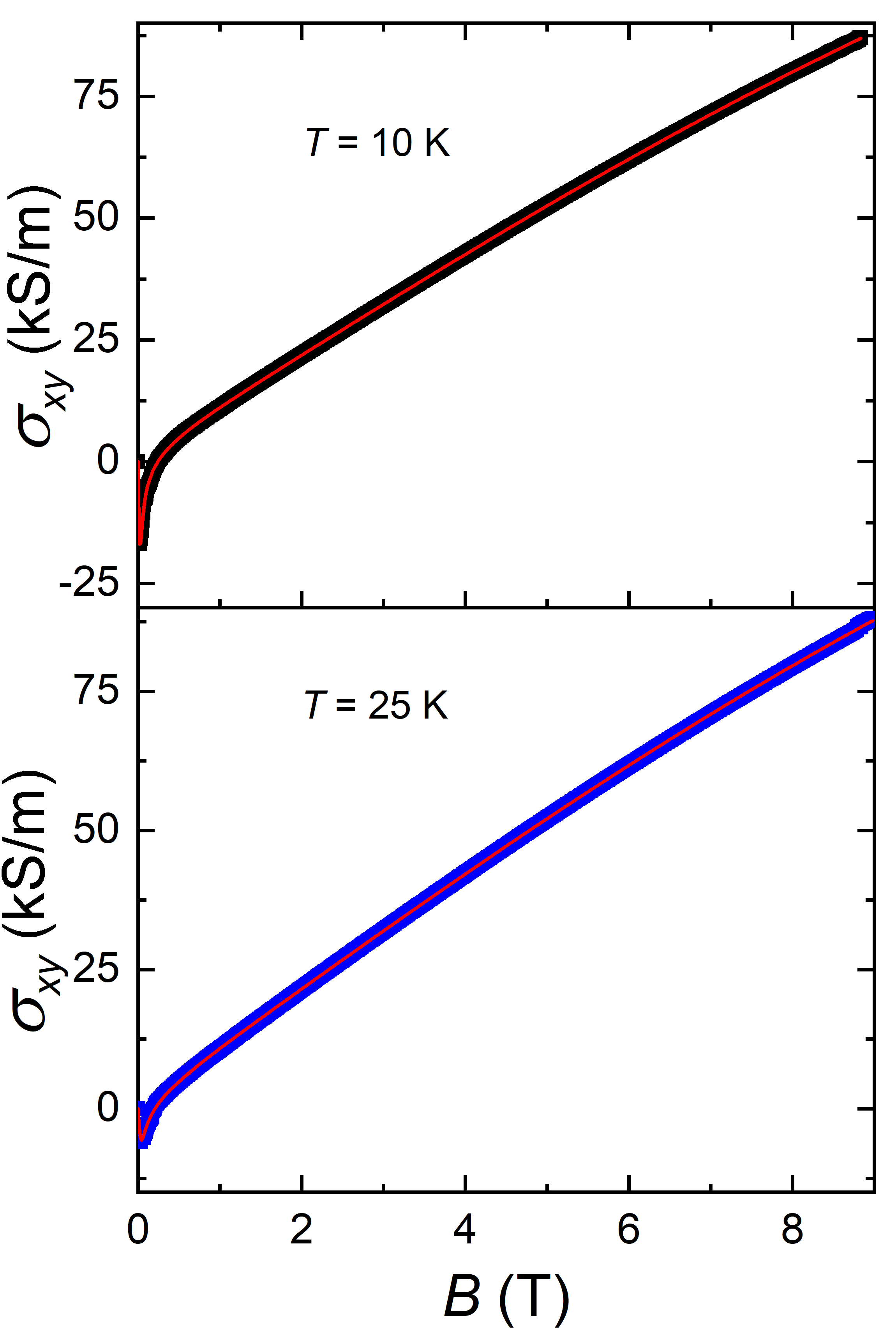}
	\includegraphics[width=0.32\textwidth]{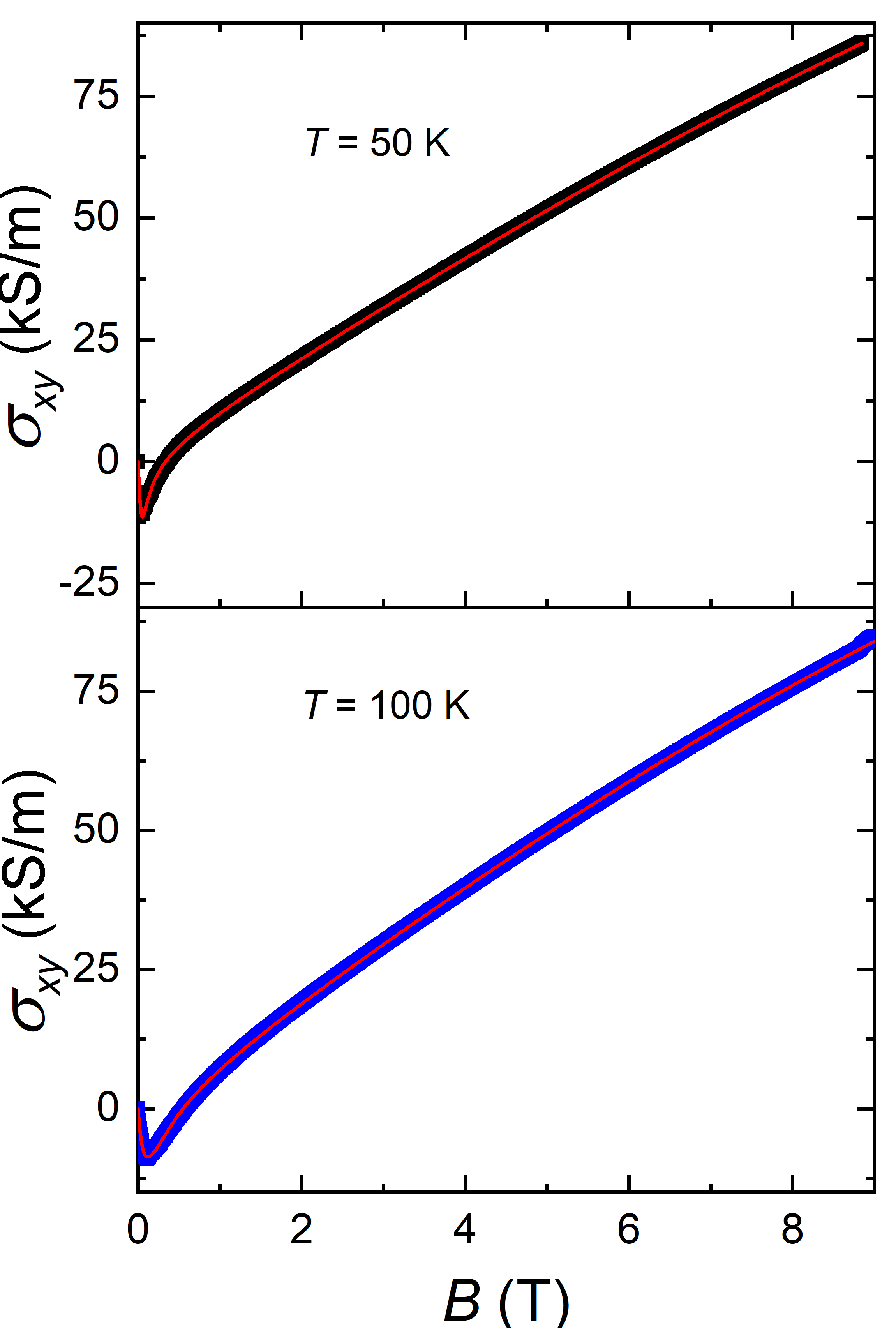}
	\includegraphics[width=0.32\textwidth]{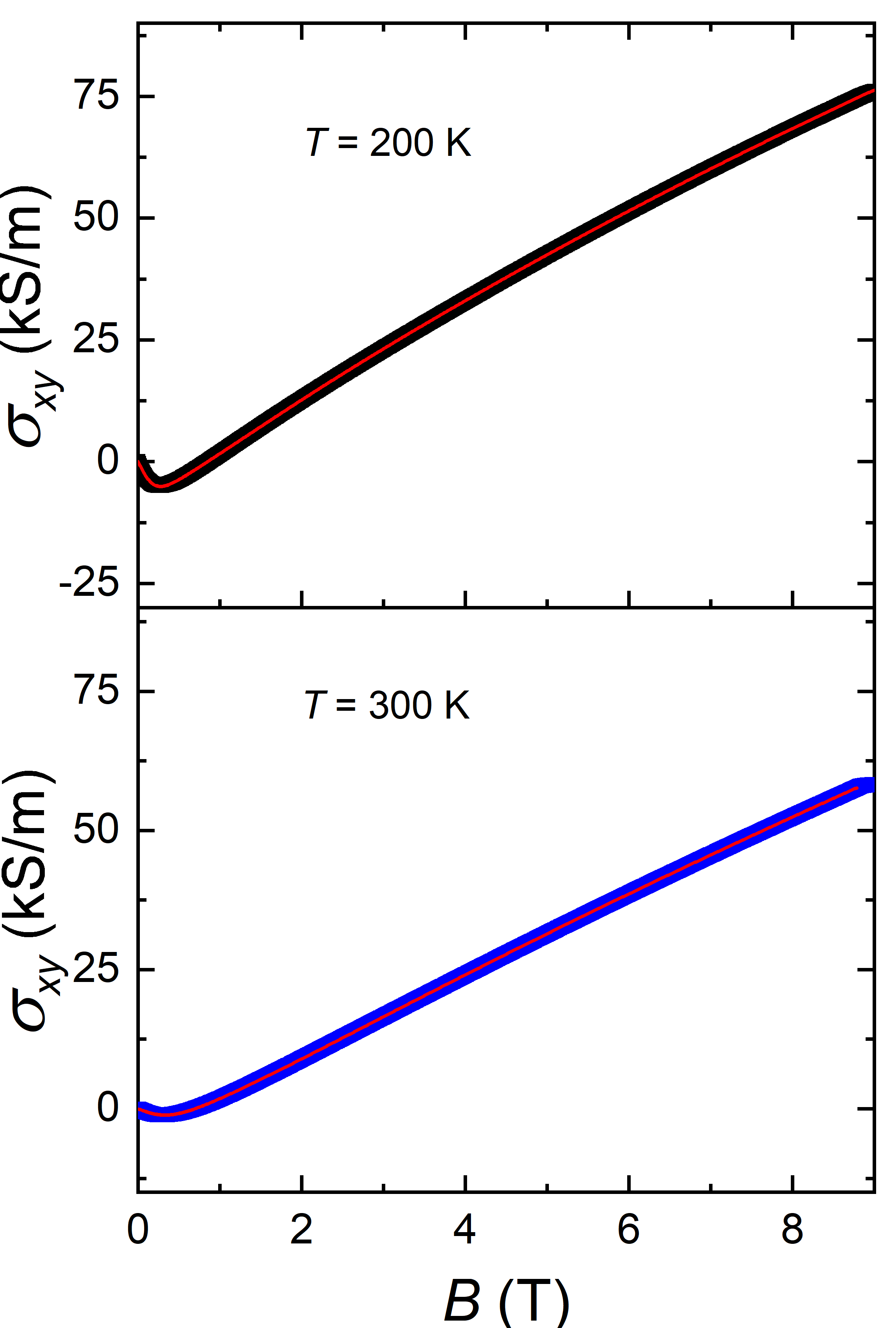}\vspace{-0.5cm}
	\caption{Hall conductivity as a function of magnetic field for several temperatures. 
	Red solid lines are the fits of the three-bands Drude model (Eq.~S1). Parameters of the fitted models are collected in Table~S2. Quality of all fits was excellent (with $R^2>$99.99).}
	\label{Hall_img}
\end{figure*}

For a quantitative analysis of our data we used three-band Drude model of Hall conductivity (Eq.~1 of the main text):
\begin{equation}
\sigma_{xy}(B)=\sum_{i=1}^3\frac{e\,n_i \mu_i^2B}{1+(\mu_iB)^2},
\label{three-band-sigmaxy}
\end{equation}
with $n_i$ and $\mu_i$ denoting respectively concentration and
mobility of carriers from the $i$-th band ($e$ is the elementary charge).
We fitted Eq.~\ref{three-band-sigmaxy} to the experimental Hall conductivity data: $\sigma_{xy}=\rho_{xy}/(\rho_{xx}^2+\rho_{xy}^2)$ (because $\rho_{xy}\!\ll\!\rho_{xx}$).
The results of this fitting to $\sigma_{xy}(B)$ obtained at temperatures higher than 2~K are shown  in Fig.~\ref{Hall_img} (as red solid lines). Parameters of all fits and collected in Table~S2. Interestingly, our analysis demonstrates that between $T=\!50$\,K and $T=\!100$\,K hole-like character of one of bands changes to electron-like. This behavior can possibly be due to a temperature-driven Lifshitz transition, previously observed in topological semimetals, for example  WTe$_2$  [10] and ZrTe$_5$ [11,12]. 
\begin{table*}[b]
	\raggedright{\textbf{TABLE S2} Parameters obtained for ScPtBi from three-bands Drude model fitting to Hall conductivity shown in Fig.~1 (main text) and Fig.~\ref{Hall_img}: $n_i$ are carrier concentrations, $\mu_i$ are carrier mobilities (for $i=1,2,3$). Carriers with positive $n_i$ are holes, those with negative $n_i$ are electrons. All fit parameters were obtained with errors smaller than 7\%.  }\\\vspace{0.4cm}
	\begin{tabular*}{0.95\textwidth}{@{\extracolsep{\fill}}*{7}{c}} \hline\hline
		$T$ & $n_1$ & $\mu_1$& $n_2$ & $\mu_2$ & $n_3$ & $\mu_3$ \\
		 (K) & ($\rm{cm^{-3}}$) & ($\rm{cm^2V^{-1}s^{-1}}$) & ($\rm{cm^{-3}}$) & ($\rm{cm^2V^{-1}s^{-1}}$) & ($\rm{cm^{-3}}$) & ($\rm{cm^2V^{-1}s^{-1}}$) \\\hline
		2 & $5.47\!\times\!10^{19}$ & $351$ & $-4.55\!\times\!10^{15}$ & $212816$ & $1.14\!\times\!10^{16}$ & $10386$\\
		10 & $5.33\!\times\!10^{19}$ & $356$ & $-4.98\!\times\!10^{15}$ & $435109$ & $1.11\!\times\!10^{16}$ & $12710$\\
		25 & $5.51\!\times\!10^{19}$ & $348$ & $-3.34\!\times\!10^{15}$ & $229137$ & $6.91\!\times\!10^{15}$ & $11706$\\
		50 & $5.59\!\times\!10^{19}$ & $343$ & $-8.34\!\times\!10^{15}$ & $179416$ & $1.50\!\times\!10^{16}$ & $6105$\\
		100 & $5.04\!\times\!10^{19}$ & $358$ & $-2.87\!\times\!10^{15}$ & $166572$ & $-1.90\!\times\!10^{16}$ & $50797$\\
		200 & $6.87\!\times\!10^{19}$ & $290$ & $-2.49\!\times\!10^{16}$ & $30429$ & $-6.87\!\times\!10^{16}$ & $7577$\\
		300 & $1.18\!\times\!10^{20}$ & $192$ & $-2.25\!\times\!10^{16}$ & $16250$ & $-9.14\!\times\!10^{16}$ & $4591$\\
		\hline\hline
	\end{tabular*}\label{Hall_parameters}
\end{table*}
\clearpage
\subsection*{Weak antilocalization analysis and CMA parameters for different temperatures}
\begin{figure*}[h]
	\includegraphics[width=0.45\textwidth]{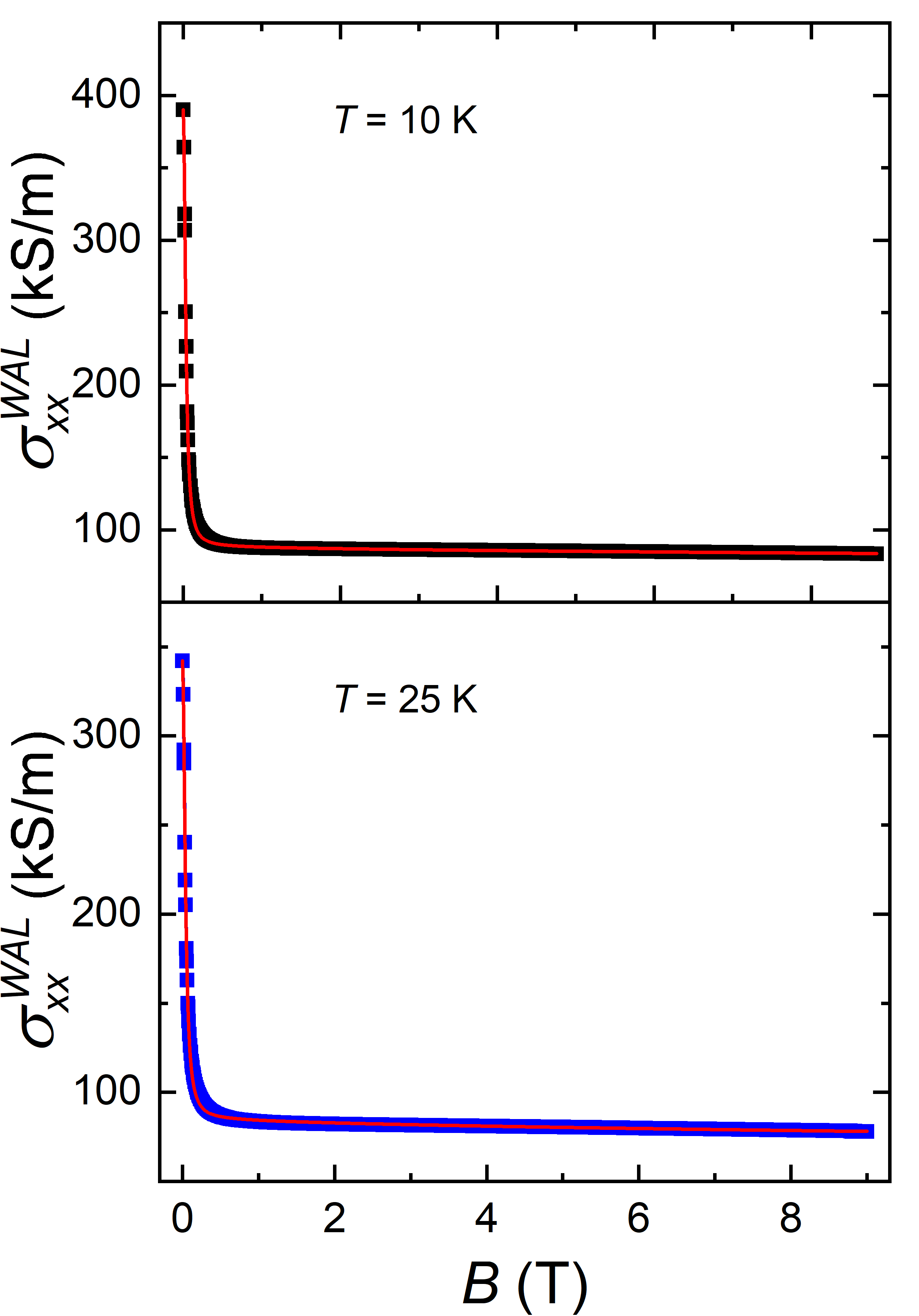}
	\includegraphics[width=0.45\textwidth]{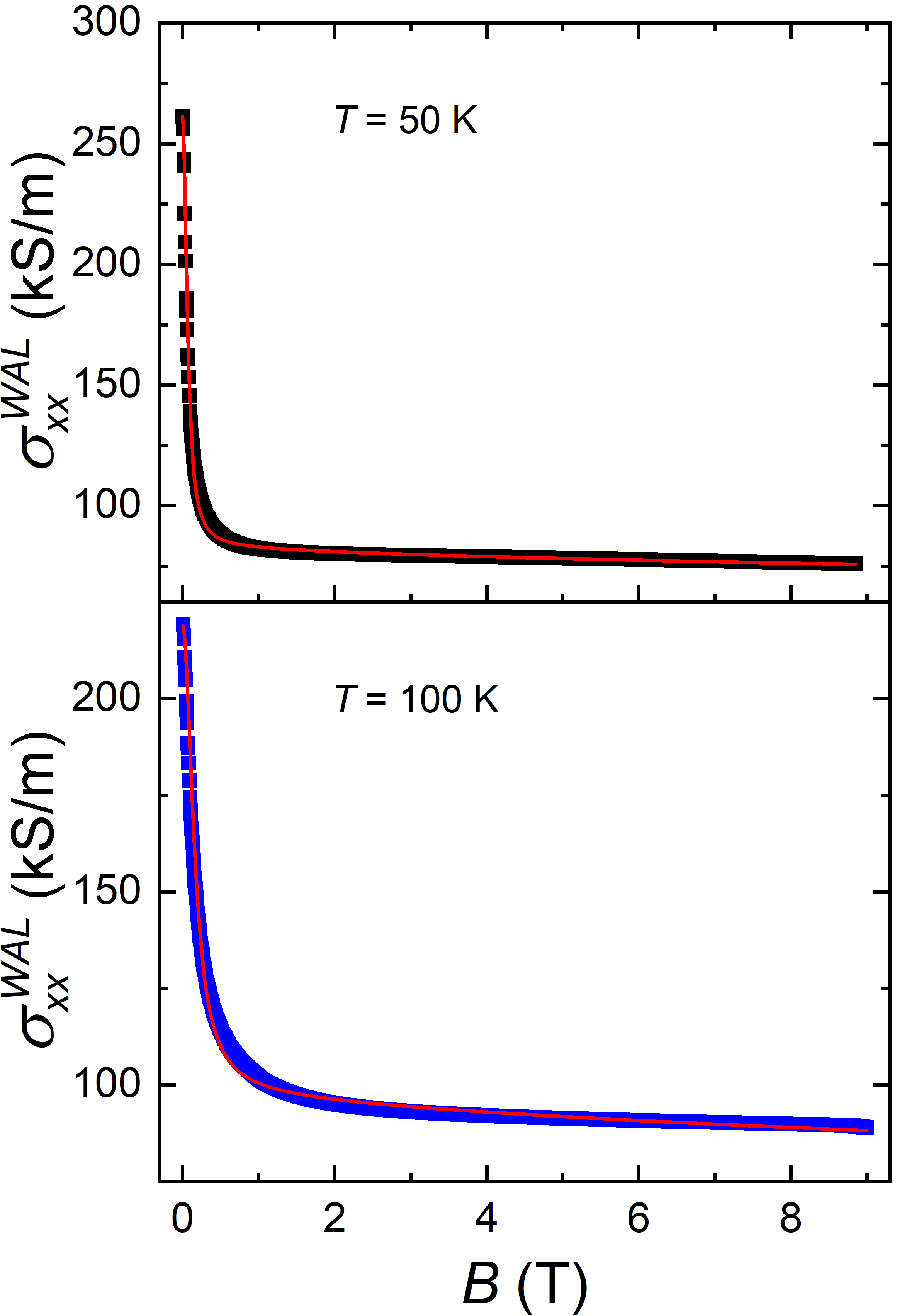}\\
	\includegraphics[width=0.45\textwidth]{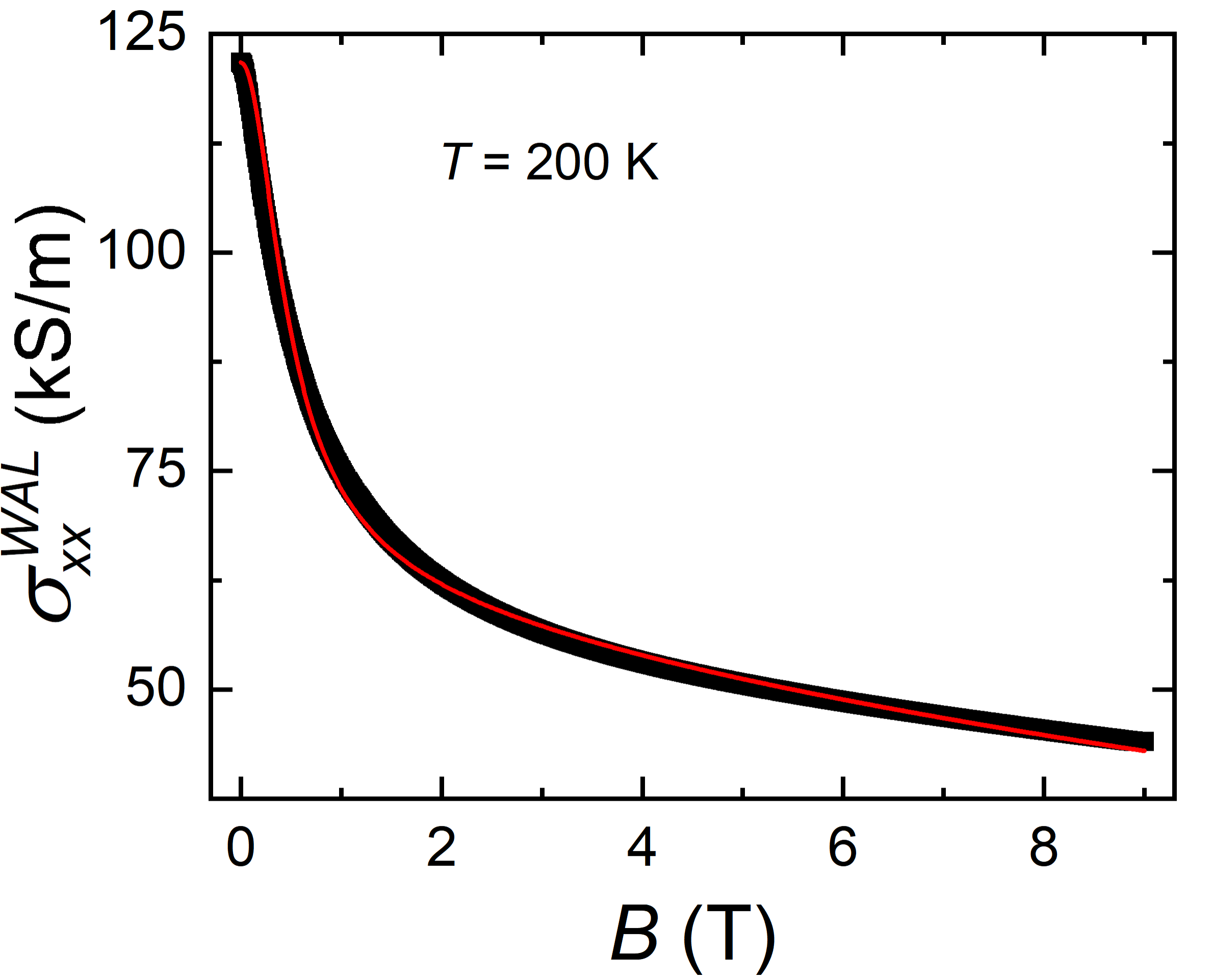}
	\caption{Extracted contribution of weak antilocalization to longitudinal electrical conductivity of ScPtBi as a function of magnetic field, for several temperatures. Red  lines correspond to the fits with Eq.~3 of the main text. Parameters of the fitted models are collected in Table~S3A.}
	\label{WAL_img}
\end{figure*}
\clearpage
\begin{table*}[h]
	\raggedright{\textbf {TABLE S3} Weak antilocalization and chiral magnetic anomaly parameters
	for longitudinal electrical conductivity of ScPtBi at different temperatures}\\\vspace{0.5cm}
	\begin{tabular*}{\textwidth}{@{\extracolsep{\fill}}*{6}{c}}\hline\hline 
	  \multicolumn{6}{l}{{\bf A} -- WAL parameters obtained from fits shown in Fig.~2C (main text) and Fig.~\ref{WAL_img}}\\
	\hline
		$T$ & $B_{c}$ & $l_{\phi}$& $C_1^{qi}$ & $C_2^{qi}$ & $\sigma_0$ \\
		(K) & (T) & (nm) & (S\,m$^{-1}$\,T$^{-0.5}$) & (S\,m$^{-1}$\,T$^{-2}$) & (S\,m$^{-1}$) \\\hline
		2 & $0.02899$ & $150.65$ & $-2369$ & $-4.58\!\times\!10^8$ & $4.72\!\times\!10^5$\\
		10 & $0.03379$ & $139.53$ & $-2085$ & $-2.62\!\times\!10^8$ & $3.9\!\times\!10^5$\\
		25 & $0.03968$ & $128.77$ & $-2947$ & $-1.62\!\times\!10^8$ & $3.42\!\times\!10^5$\\
		50 & $0.06713$ & $98.99$ & $-3263$ & $-3.9\!\times\!10^7$ & $2.61\!\times\!10^5$\\
		100 & $0.16049$ & $64.03$ & $-4651$ & $-4.55\!\times\!10^6$ & $2.19\!\times\!10^5$\\
		200 & $0.44143$ & $38.61$ & $-10317$ & $-2.46\!\times\!10^5$ & $1.22\!\times\!10^5$\\
		\hline
	\end{tabular*}
	\begin{tabular*}{\textwidth}{l c c c c c c c} \hline 
	\multicolumn{8}{l}{{\bf B} -- WAL and CMA parameters obtained from fits of Eq.~3, as shown in Fig. 2D (main text)}\\
	\hline
		$T$ & $B_{c}$ & $l_{\phi}$& $C_1^{qi}$ & $C_2^{qi}$ & $\sigma_0$ & $\sigma_n$ & $C_W$ \\
		(K) & (T) & (nm) & (S\,m$^{-1}$\,T$^{-0.5}$) & (S\,m$^{-1}$\,T$^{-2}$) & (S\,m$^{-1}$) & (S\,m$^{-1}$) & (T$^{-2}$)\\\hline
		2 &~~~~~$0.08621$~~~~~& $87.36$ & $-45434$ & $-3.3\!\times\!10^7$ & $4.72\!\times\!10^5$ & $3.25\!\times\!10^5$ & $0.0137$\\
		25 & $0.07837$ &~~~$91.62$~~~~& $-27675$ & $-3.20\!\times\!10^7$ & $3.42\!\times\!10^5$ & $3.21\!\times\!10^5$ & $0.00906$\\
		50 & $0.11497$ & $75.65$ & $-15474$ & $-1.12\!\times\!10^7$ &~~~$2.61\!\times\!10^5$~~~& $3.33\!\times\!10^5$ & $0.00275$\\
		100~~~& $0.20778$ & $56.27$ & $-14966$ & $-2.1\!\times\!10^6$ & $2.19\!\times\!10^5$ &~~~$3.12\!\times\!10^5$~~~&~~~$0.00137$~~~\\
		\hline\hline
	\end{tabular*}\label{WAL_CMA_parameters}
\\
\flushleft WAL parameters obtained from analysis of TMR (A) and LMR (B) are somewhat different, however similar differences were reported for a Dirac semimetal Bi$_{0.97}$Sb$_{0.03}$ by Kim et al. [13].
\end{table*}
Because the conductivity related to CMA  (Eq. 17 in [14]) is: $\frac{e^4}{4\pi^2 \hbar c^2} \frac{v^3\tau}{\mu_c^2}B^2$ we may consider $C_W$ proportional to $\frac{v^3\tau}{\mu_c^2}$ (where $v$ is Fermi velocity, $\tau$ intervalley scattering relaxation time, and $\mu_c$  is chemical potential measured from Weyl point). Small value of $C_W$ may thus implicate small $v$ and/or $\tau$, and/or large $\mu_c$. Large mobilities obtained from Hall effect analysis indicate large $v$ and $\tau$, therefore the implication of a small $C_W$ in ScPtBi is most likely the large value of $\mu_c$. 
\clearpage
\subsection*{Supplementary results of electronic structure calculations}
\begin{figure*}[h]
	\includegraphics[width=0.55\textwidth]{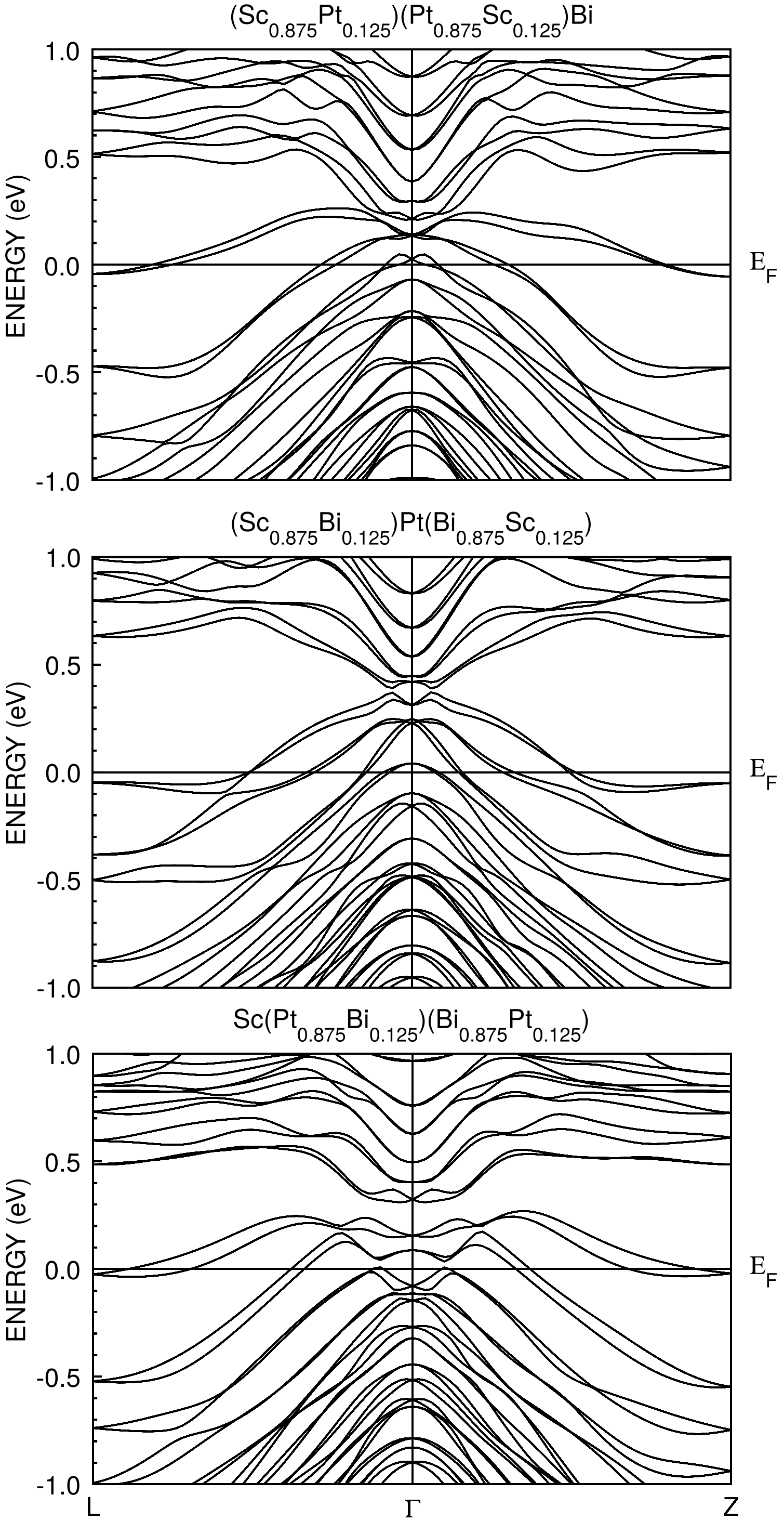}
	\caption{Electronic band structure of ScPtBi for crystal structure models with different antisite defects. Note both hole-like and electron-like bands crossing the Fermi energy $E_{\rm F}$. 
	\label{electronic_structure}}
\end{figure*}

\begin{figure*}[b]
	\includegraphics[width=0.43\textwidth]{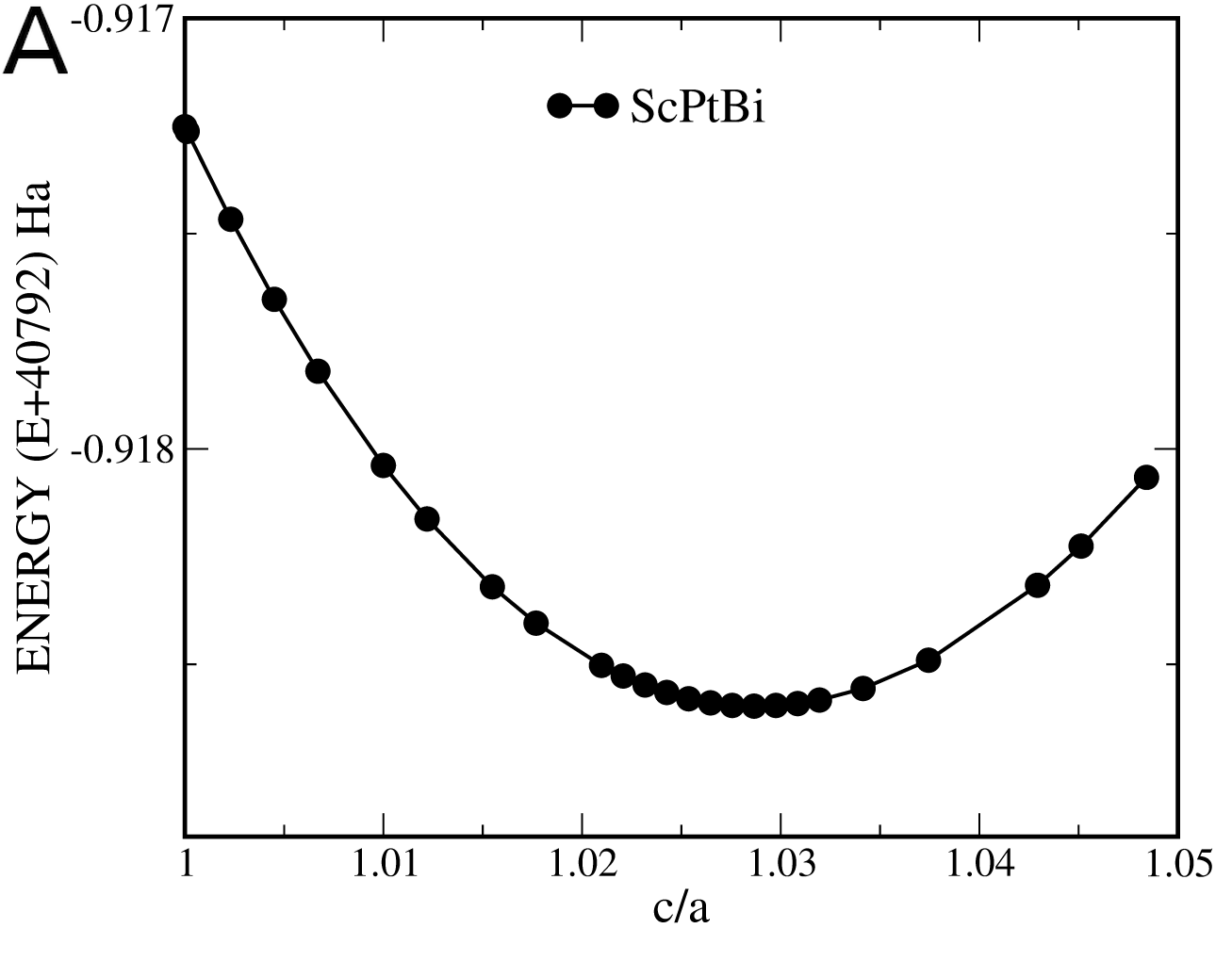}
	\includegraphics[width=0.54\textwidth]{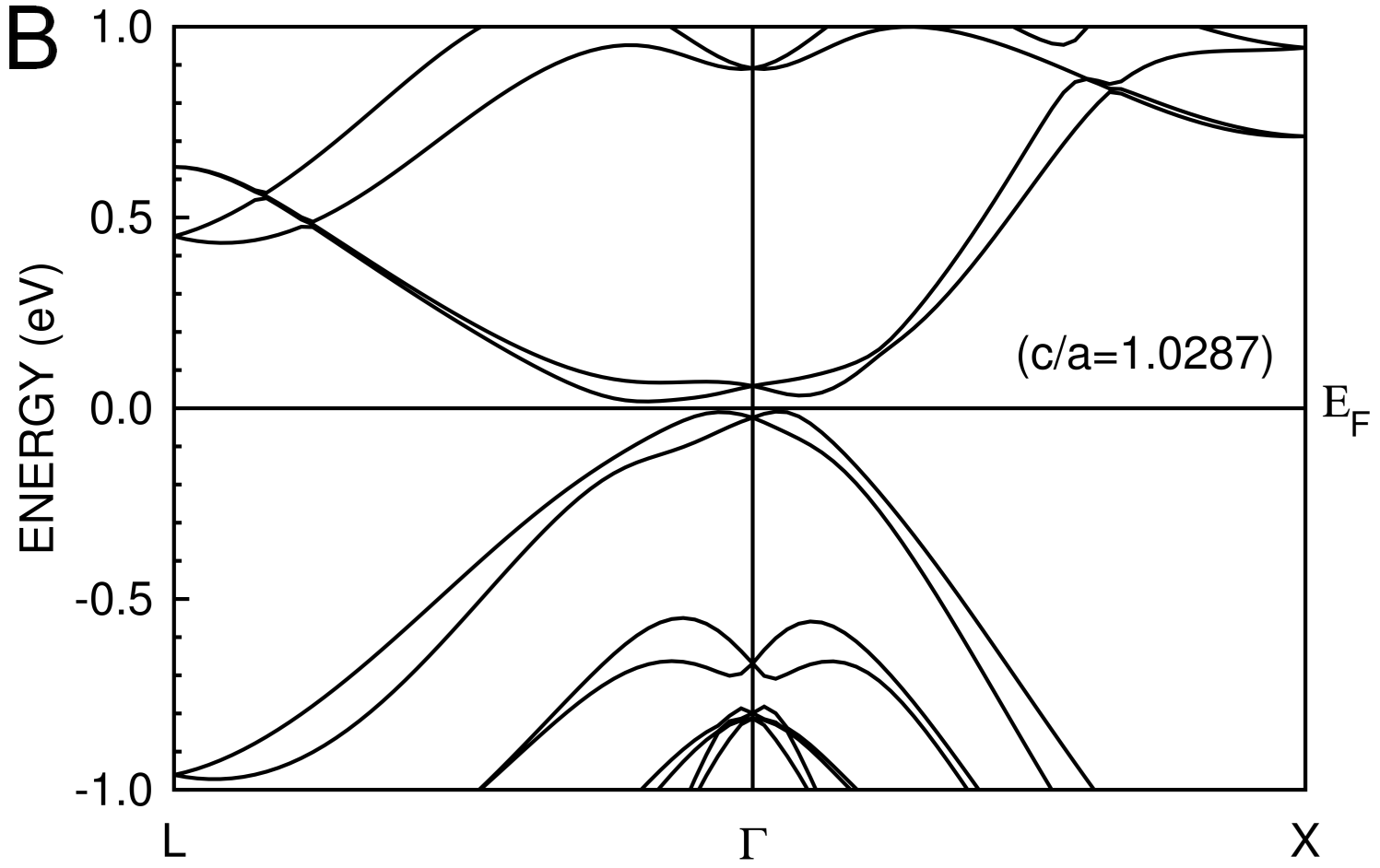}\\
	\includegraphics[width=0.48\textwidth]{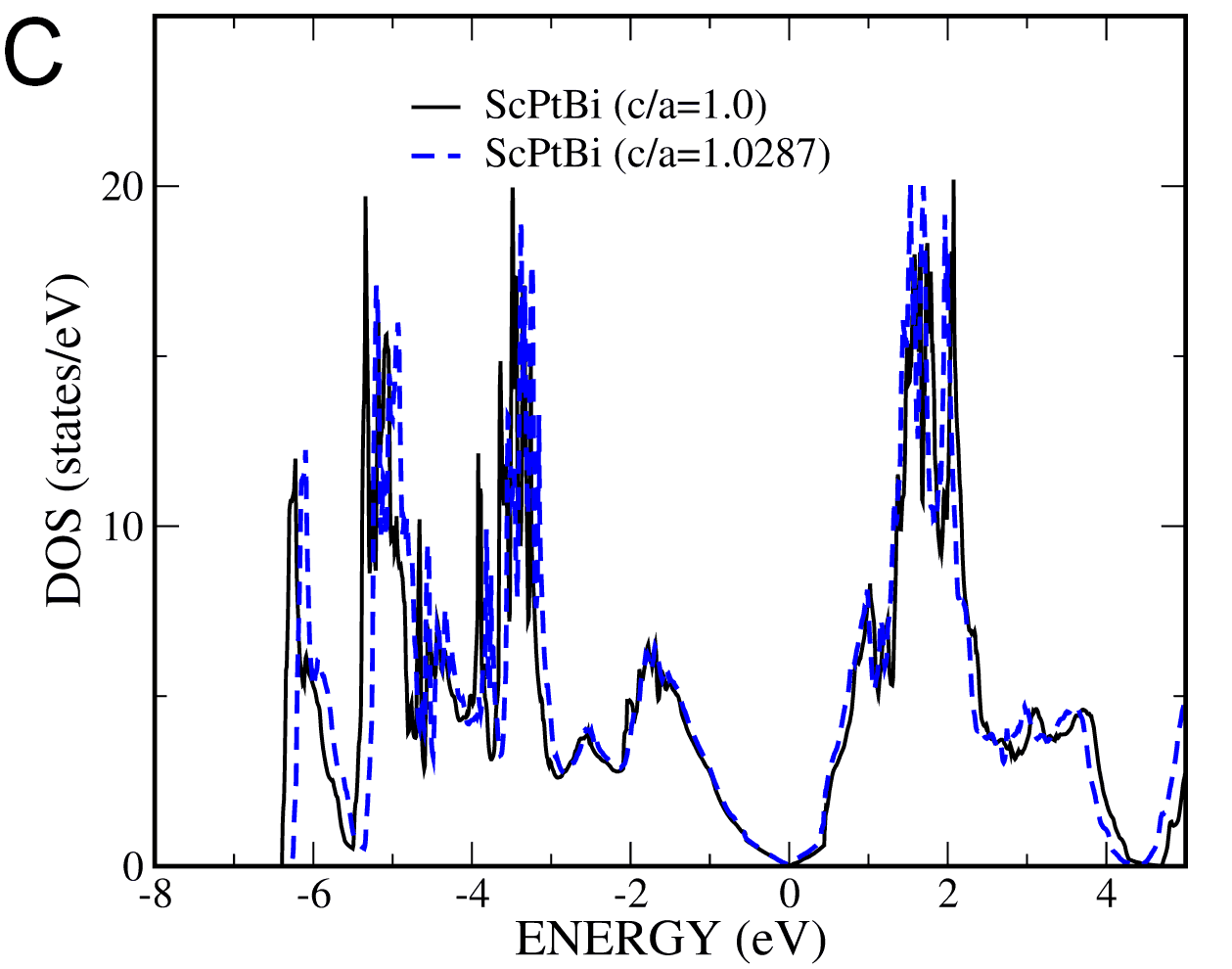}
	\includegraphics[width=0.48\textwidth]{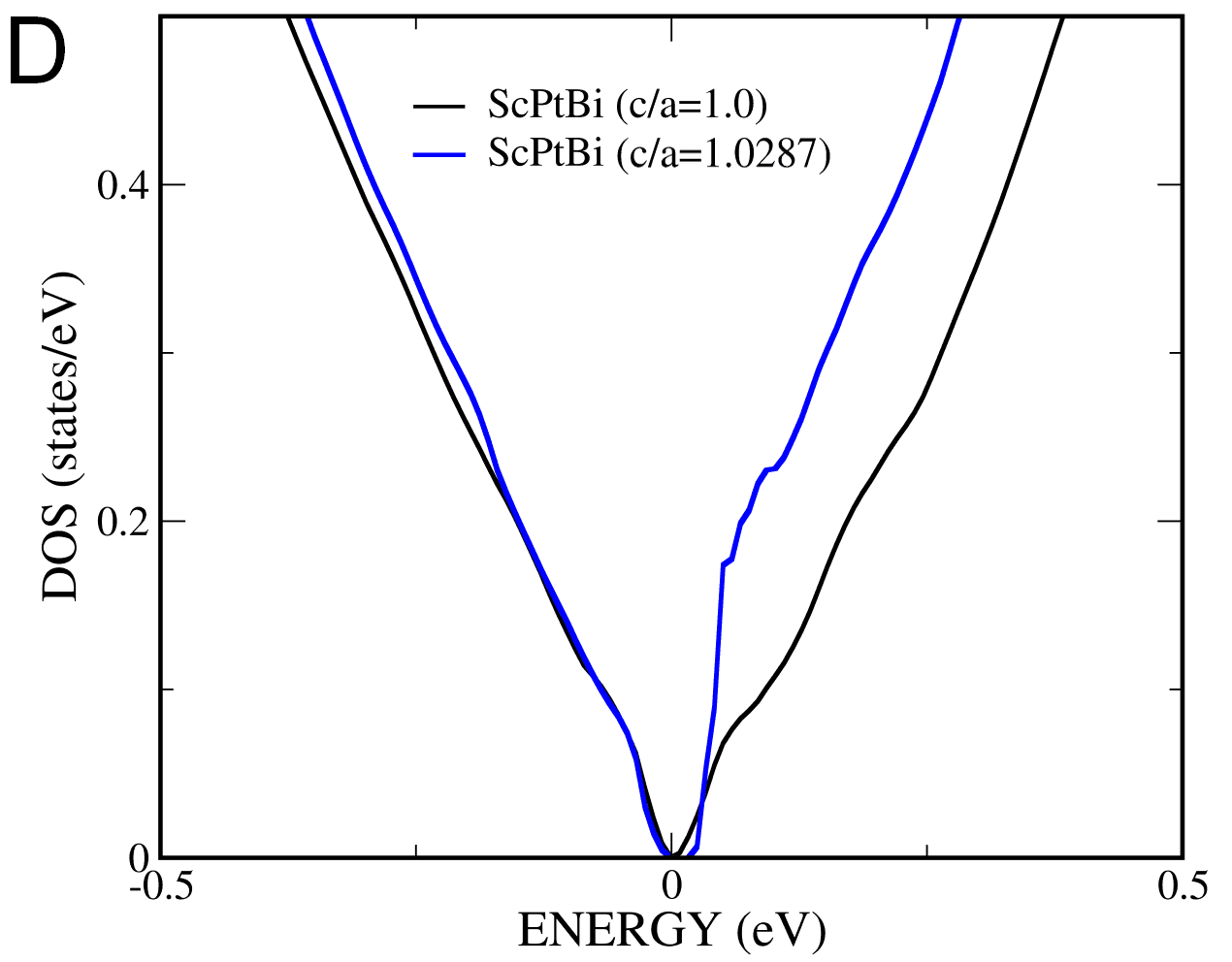}
	\caption{(A) Total energy calculated for tetragonally distorted crystal structure of ScPtBi  as a function of ratio of lattice parameters, $c/a$. (B) Electronic structure of tetragonally distorted ScPtBi, with $c/a = 1.0287 $ - corresponding to energy minimum in panel (A). (C) Calculated density of states in tetragonally distorted ScPtBi (with $c/a = 1.0287$) - blue dashed line, compared with density of states of ScPtBi with cubic crystal structure - black solid line. (D) Same as in (C) but in close vicinity of Fermi level. For $c/a = 1.0287$ DOS$(E_{\rm F})$ is zero, and there is a gap of 23~meV.
	\label{el_struct_tetradistor}}
\end{figure*}

\vspace{-0.6cm}
We show in Fig.~\ref{electronic_structure} band structures for models with different antisite defects, and in Fig.~\ref{el_struct_tetradistor} total energy, band structure and DOS calculated for the model of tetragonally distorted ScPtBi.
 \clearpage
For the model of tetragonally distorted ScPtBi we obtained results shown in Fig.~\ref{el_struct_tetradistor}. First, we calculated  total energy for crystal structure of ScPtBi  as a function of ratio of lattice parameters, $c/a$ (Fig.~\ref{el_struct_tetradistor}A). The lowest energy was found for $c/a = 1.0287$, and for such a distortion we calculated band structure and DOS, as shown in Fig.~\ref{el_struct_tetradistor}B-D, revealing a gap of 23 meV. 
It closely resembles results presented by Ding, Gao, Yu, Ni, and Yao [15], as well as by Kaur, Dhiman, and Kumar [16]. In these both papers the increase of the $c/a$ ratio by 5\% induced opening of the gap of $\approx70$\,meV, however in the latter paper, for a 3\% distortion the gap remained closed, most likely due to different calculation codes applied. \vspace{2cm}
\clearpage
\noindent {\bf Supplemental references:}\\
\noindent [1]	 K. Synoradzki, K. Ciesielski, I. Veremchuk, H. Borrmann, P. Skokowski, D. Szymański, Y. Grin, and D. Kaczorowski, Materials. {\bf 12}, 1723 (2019).\\
\noindent [2]	 Y. Nakajima, R. Hu, K. Kirshenbaum, A. Hughes, P. Syers, X. Wang, K. Wang, R. Wang, S. R. Saha, D. Pratt, J. W. Lynn, and J. Paglione, Sci. Adv. {\bf 1}, e1500242 (2015).\\
\noindent [3]	 O. Pavlosiuk, D. Kaczorowski, and P. Wiśniewski, Sci. Rep. {\bf 5}, 9158 (2015).\\
\noindent [4] O. Pavlosiuk, D. Kaczorowski, X. Fabreges, A. Gukasov, and P. Wiśniewski, Sci. Rep. {\bf 6}, 18797 (2016).\\
\noindent [5]	 E. D. Mun, S. L. Bud’ko, C. Martin, H. Kim, M. A. Tanatar, J.-H. Park, T. Murphy, G. M. Schmiedeshoff, N. Dilley, R. Prozorov, and P. C. Canfield, Phys. Rev. B {\bf 87}, 075120 (2013).\\
\noindent [6]	 O. Pavlosiuk, D. Kaczorowski, and P. Wiśniewski, Phys. Rev. B {\bf 94}, 035130 (2016).\\
\noindent [7]	 N. P. Butch, P. Syers, K. Kirshenbaum, A. P. Hope, and J. Paglione, Phys. Rev. B {\bf 84}, 220504 (2011).\\
\noindent [8]	Z. Hou, Y. Wang, E. Liu, H. Zhang, W. Wang, and G. Wu, Appl. Phys. Lett. {\bf 107}, 202103 (2015).\\
\noindent [9]	M. Hirschberger, S. Kushwaha, Z. Wang, Q. Gibson, S. Liang, C. A. Belvin, B. A. Bernevig, R. J. Cava, and N. P. Ong, Nat. Mater. {\bf 15}, 1161 (2016).\\
\noindent [10] Y. Wu, N. H. Jo, M. Ochi, L. Huang, D. Mou, S. L. Bud’ko, P. C. Canfield, N. Trivedi, R. Arita, and A. Kaminski, Phys. Rev. Lett. {\bf 115}, 166602 (2015).\\
\noindent [11] Y. Zhang, C. Wang, L. Yu, G. Liu, A. Liang, J. Huang, S. Nie, X. Sun, Y. Zhang, B. Shen, J. Liu, H. Weng, L. Zhao, G. Chen, X. Jia, C. Hu, Y. Ding, W. Zhao, Q. Gao, C. Li, S. He, L. Zhao, F. Zhang, S. Zhang, F. Yang, Z. Wang, Q. Peng, X. Dai, Z. Fang, Z. Xu, C. Chen, and X. J. Zhou, Nat. Commun. {\bf 8}, 15512 (2017).\\
\noindent [12] H. Chi, C. Zhang, G. Gu, D.E. Kharzeev, X. Dai, and Q. Li, New J. Phys. {\bf 19}, 015005 (2017).\\
\noindent [13] H.-J. Kim, K.-S. Kim, J.-F. Wang, M. Sasaki, N. Satoh,  A. Ohnishi, M. Kitaura, M. Yang, and L. Li, Phys. Rev. Lett. {\bf 111}, 246603 (2013).\\
\noindent [14] D.~T. Son and B.~Z. Spivak, Phys. Rev. B {\bf 88}, 104412 (2013).\\
\noindent [15] G. Ding, G.~Y. Gao, L. Yu, Y. Ni, and K. Yao, J. Appl. Phys. {\bf 119}, 025105 (2016).\\
\noindent [16] K. Kaur, S. Dhiman, and R. Kumar, Phys. Lett. A {\bf 381}, 339 (2017). 
\end{document}